\documentclass[orivec]{llncs}

	\let\llncssubparagraph\subparagraph
	\let\subparagraph\paragraph
\input package-and-newcommand-includes

	\let\subparagraph\llncssubparagraph
\RequirePackage[top=2.0cm,bottom=2.0cm,left=3.8cm,right=3.8cm,nohead]{geometry}

\usepackage[title,toc,titletoc,header]{appendix}

\begin{document}

\let\jnote=\ignore
\let\tnote=\ignore
\let\tsumm=\ignore

\pagestyle{plain}

\title{OIM: {Oscillator-based Ising Machines for Solving Combinatorial Optimisation Problems}}
  
\author{Tianshi Wang \and Jaijeet Roychowdhury}
\authorrunning{T. Wang and J. Roychowdhury}
\institute{
Department of Electrical Engineering and Computer Sciences, University of California, Berkeley \\
\email{\{tianshi,jr\}@berkeley.edu}}

\maketitle

\begin{abstract}
We present a new way to make Ising machines, \ie, using networks of coupled self-sustaining nonlinear oscillators. Our scheme is theoretically rooted in a novel result that establishes that the phase dynamics of coupled oscillator systems, under the influence of sub-harmonic injection locking, are governed by a Lyapunov function that is closely related to the Ising Hamiltonian of the coupling graph. As a result, the dynamics of such oscillator networks evolve naturally to local minima of the Lyapunov function. Two simple additional steps (\ie, adding noise, and turning sub-harmonic locking on and off smoothly) enable the network to find excellent solutions of Ising problems. We demonstrate our method on Ising versions of the MAX-CUT and graph colouring problems, showing that it improves on previously published results on several problems in the G benchmark set. Our scheme, which is amenable to realisation using many kinds of oscillators from different physical domains, is particularly well suited for CMOS IC implementation, offering significant practical advantages over previous techniques for making Ising machines. We present working hardware prototypes using CMOS electronic oscillators.
\end{abstract}

\thispagestyle{plain}

\section{Introduction}\seclabel{intro}

% Ising model/problem/machines and connection with combinatorial optimization problems

The Ising model \cite{ising1925beitrag,Brush1967RevModPhysHistoryOfIsingModel} takes any weighted graph and uses it to define a scalar function called the Ising Hamiltonian. 
Each vertex in the graph is associated with a \textit{spin}, \ie, a binary variable taking values $\pm 1$. 
The Ising problem is to find an assignment of spins that minimises the Ising Hamiltonian (which depends on the spins and on the graph's weights). 
Solving the Ising problem in general has been shown to be very difficult \cite{barahona1982computational}, but devices that can solve it quickly using specialised hardware have been proposed in recent years \cite{marandi2014network,mcmahon2016ScienceIsing100,inagaki2016ScienceIsing2000,johnson2011quantum,bian2014Ising,yamaoka2016IsingCMOS}.
Such Ising machines have attracted much interest because many classically difficult combinatorial optimisation problems (including all 21 of Karp's well-known list of NP-complete problems \cite{karp1972np}) can be mapped to Ising problems \cite{lucas2013ising}. 
Hence, as Moore's Law nears its limits, Ising machines offer promise as a novel alternative paradigm for solving difficult computational problems effectively.

% short 1 para summary of what we do here
We present a new and attractive means for realising Ising machines, \ie, using networks of coupled, self-sustaining nonlinear oscillators.
We first establish a key theoretical result that relates the (continuous) phase dynamics of an oscillator network with the (discrete/combinatorial) Ising Hamiltonian of the graph representing the oscillator couplings.
We then build on this result to develop practical oscillator-based Ising machines and demonstrate effectiveness by solving the NP-complete MAX-CUT and graph colouring combinatorial optimisation problems \cite{festa2002randomized,Jensen2011GraphColoringProblems}. 
We present working hardware prototypes of our oscillator based Ising machines.

% expand a bit on the Lyapunov function and the Ising Hamiltonian

We first show that the phase dynamics of any network of coupled, self-sustaining, amplitude-stable oscillators can be abstracted using the Generalised Adler model \cite{NeRoDATE2012SHIL,BhRoASPDAC2009}, a generalisation of the well-known Kuramoto model \cite{Kuramoto1975,Kuramoto2003,acebron2005kuramoto}. \ignore{or Kuramoto-Sakaguchi \cite{KS} - KS still uses sin(), but does sin(theta_i-theta_j+alpha)} 
The model's phase dynamics are governed by an associated Lyapunov function, \ie, a scalar function of the oscillators' phases that is always non-increasing and settles to stable local minima as phase dynamics evolve. 
If each oscillator's phase settles to either 0 or $\pi$ (radians) and these values are associated with spins of $\pm 1$, we show that this Lyapunov function is essentially identical to the Ising Hamiltonian of the oscillator network's connectivity graph. 
In general, however, oscillator phases do not settle to the discrete values 0/$\pi$, but span a continuum of values instead. 
In order to binarise oscillator phases (\ie, get them to settle to values near 0/$\pi$), we inject each oscillator with a second harmonic signal (dubbed SYNC) that induces sub-harmonic injection locking (SHIL), which makes the phase of each oscillator settle to a value near either 0 or $\pi$ \cite{WaRoUCNC2014PHLOGON,NeRoDATE2012SHIL}. 
We devise a new Lyapunov function that governs the network's dynamics with SHIL; this Lyapunov function is also essentially identically to the Ising Hamiltonian at phase values of 0/$\pi$.

%\redHL{should write about the variability result, somehow}
%\redHL{using 2d and 3d KCIH plots, we should demonstrate that the minima fall at 0/pi?}

% building on this result -- noise, SYNC ramping; results
Thus we show that when SHIL binarisation is applied, coupled oscillator network dynamics settle naturally to \textit{local} minima of a continuised version of the associated Ising Hamiltonian. 
To evolve the system out of local minima towards the global minimum, we show that a simple scheme, in which the binarising second-harmonic SYNC signal's amplitude is ramped up and down and judicious amounts of noise added, works well. 
We present simulation results on a standard MAX-CUT benchmark set of 54 large problems, demonstrating not only that it finds the best-known previous results in many cases, but finds \textbf{better results than seem to have been previously published for 17 of the 54 problems}.
We also demonstrate our method on the graph-colouring problem and present small (up to 32 CMOS oscillators) prototypes built on breadboard that function perfectly, testifying to the ease with which practical hardware implementations can be built.

% mention of previous approaches and comparison
Our scheme is different from previous Ising machine approaches, which are of 3 types (see \secref{IsingIntro}): 1) a fiber-optic laser-based scheme known as the Coherent Ising Machine \cite{marandi2014network,mcmahon2016ScienceIsing100,inagaki2016ScienceIsing2000}, 2) the D-WAVE quantum Ising machine \cite{johnson2011quantum,bian2014Ising} and 3) CMOS hardware accelerated simulated annealing chips for solving Ising problems \cite{yamaoka2016IsingCMOS,AramonEtAlarXiv2018DigitalAnnealer,GyHiSaIEICE2018isingNoNoise,GyHiSaICCAD2018parallelTemperingForIsing}.
Unlike CIM and D-WAVE, which are large, expensive and ill-suited to low-cost mass production, our approach is a purely classical scheme that does not rely on quantum phenomena or novel nano-devices. 
Indeed, it can be implemented using conventional CMOS electronics, which has many advantages: scalability/miniaturisability (\ie, very large numbers of spins in a physically small system), well-established design processes and tools that essentially guarantee first-time working hardware, very low power operation, seamless integration with control and I/O logic, easy programmability via standard interfaces like USB, and low cost mass production. 
CMOS implementations of our scheme also allow complete flexibility in introducing controlled noise and programming SYNC ramping schedules. 
Furthermore, implementing oscillator coupling by physical connectivity makes our scheme inherently parallel, unlike CIM, where coupling is implemented via FPGA-based digital computation and is inherently serial. 
The advantages of CMOS also apply, of course, to hardware simulated annealing engines \cite{yamaoka2016IsingCMOS,AramonEtAlarXiv2018DigitalAnnealer,GyHiSaIEICE2018isingNoNoise,GyHiSaICCAD2018parallelTemperingForIsing}, but our scheme has additional attractive features.
One key advantage relates to variability, a significant problem in nanoscale CMOS.
For oscillator networks, device- and circuit-level variability impacts the system by causing a spread in the natural frequencies of the oscillators. 
Unlike other schemes, where performance deteriorates due to variability \cite{yamaoka2016IsingCMOS}, we can essentially eliminate variability by means of simple VCO-based calibration to bring all the oscillators to the same frequency.\footnote{Moreover, as we show in \secref{Kuramoto-variability}, our scheme is inherently resistant to variability even without such calibration.} 
Another key potential advantage stems from the continuous/analog nature of our scheme (as opposed to purely digital simulated annealing schemes). 
Computational experiments indicate that the time our scheme takes to find good solutions of the Ising problem grows only very slowly with respect to the number of spins.
\ignore{and degree of the connectivity. Tianshi: not shown in this paper}
This is a significant potential advantage over simulated annealing schemes \cite{AramonEtAlarXiv2018DigitalAnnealer} as hardware sizes scale up to large numbers of spins.
Note that we can use virtually any type of nonlinear oscillator (not just CMOS) to implement our scheme, including optical, MEMS, biochemical, spin torque device based, \etc, oscillators; however, CMOS seems the easiest and most advantageous implementation route at present, given the current state of technology.

\ignore{
%{\color{blue}
    Coupled oscillator networks often display intriguing synchronisation
    phenomena wherein spontaneous patterns arise.
    From the rhythmic synchronous flashing of fireflies to Huygens' clocks ticking
    in unison, from the molecular mechanism of circadian rhythms to the phase
    patterns in oscillatory neural circuits, the observation and study of
    synchronisation in coupled oscillators has a long and rich history.
    Engineers across many disciplines also take inspiration from these phenomena,
    \eg, to design high-performance radio frequency communication circuits and
    optical lasers.

    Recently, as Moore's Law nears its limits, research interest grows around
    alternative computational paradigms beyond conventional transistor
    technologies.
    Notably, several recent studies suggest that the synchronisation dynamics
    emerging from coupled oscillators can be leveraged to perform several
    computational tasks efficiently,
    such as pattern recognition \cite{Hoppensteadt2000synchronization,Porod2015physical,maffezzoni2015oscArray},
    edge detection \cite{SumanDatta2014neuro,Roy2015DPSTO},
    image segmentation \cite{wang2002dynamically}, \etc{}.
    With the advancement of today's nano-oscillator technology, these schemes are
    becoming more and more attractive [][].

    While the above exploration focuses on several specialised computational tasks
    mainly for image preprocessing, another recent study termed PHLOGON
    demonstrates general-purpose Boolean computation using oscillators
    \cite{WaRoUCNC2014PHLOGON}.
    In PHLOGON systems, Boolean logic values are encoded in the phase of
    oscillation\footnote{As an example, logic 1 can be encoded in an oscillatory
    signal aligned in phase with a reference signal; a signal with $180^\circ$
    phase shift from the reference encodes logic 0.} as opposed to static voltage
    levels.
    As it turns out, the concept of this logic encoding scheme can date back to
    decades ago.
    It was first pioneered by John von Neumann and Eiichi Goto in the 1950s
    \cite{vonNeumann:1954:NLC:nonote,Goto:1954}; computers based on it, built with
    Goto's passive resonant circuits known as Parametrons, were once commercially
    successful in Japan.
    Their scheme eventually lost its popularity due to Parametron's disadvantages
    in size and miniaturisability compared with transistors.
    PHLOGON revisits these early ideas, but instead, uses self-sustaining
    oscillators as the substrate, thus does not suffer from the same limitations.
    In fact, it offers potential advantages in energy and speed
    \cite{RoPHLOGONprocIEEE2015} compared with conventional level-based logic
    computation systems.

    A differentiating feature of PHLOGON is that it digitises the phases of
    oscillators for encoding Boolean logic values, as opposed to letting them lock
    to analog phases, \ie, continuous values between 0 and $2\pi$.
    To do so, a synchronisation signal (SYNC) oscillating at integer multiples of
    the oscillator frequency is used to excite multiple stable phase-locked
    responses in the oscillator through a mechanism known as subharmonic injection
    locking (SHIL) \cite{NeRoDATE2012SHIL,WaRoDAC2015MAPPforPHLOGON}.
    For binary logic computation, SYNC is at approximately twice the oscillator's
    natural frequency; the oscillator then develops bistable phase locks separated
    by $180^\circ$.
    Such a Booleanised oscillator can then be used as a binary logic latch in
    phase-encoded finite state machines.

    In this paper, we take this idea one step further and report a new finding ---
    a network of such Booleanised oscillators can minimise the objective function
    of virtually any combinatorial optimisation problem.
    In other words, it physically realises an Ising machine suitable for solving
    many real-world optimisation problems.

} % blue/ignore }

In the remainder of this paper, we first provide a brief summary of the Ising problem and existing Ising machine schemes in \secref{IsingIntro}.
We then present our oscillator-based Ising machine scheme (dubbed OIM, for Oscillator Ising Machine) in \secref{main}, explaining the theory that enables it to work.
Then in \secref{examples}, we present both computational and hardware examples showing the effectiveness of our scheme for solving several combinatorial optimisation problems.

\section{The Ising problem and existing Ising machine approaches}
\seclabel{IsingIntro}

The Ising model is named after the German physicist Ernest Ising.
It was first studied in the 1920s as a mathematical model for explaining domain formation in ferromagnets \cite{ising1925beitrag}.
It comprises a group of discrete variables $\{s_i\}$, \aka\/ spins, each taking a binary value $\pm 1$, such that an associated ``energy function'', known as the Ising Hamiltonian, is minimised:

\be{IsingH}
%\begin{align}
    \min ~~ H \triangleq - \sum_{1 \leq i<j \leq n} J_{ij} s_i s_j - \sum_{i=1}^n h_i s_i, %\eqnlabel{IsingH}%\\
    \quad \text{such that} ~~ s_i \in \{-1,~+1\},
%\end{align}
\ee
where $n$ is the number of spins; $\{J_{ij}\}$ and $\{h_i\}$\footnote{$\{h_i\}$ coefficients are also known as self terms in the Ising Hamiltonian.} are real coefficients.

The Ising model is often simplified by dropping the $\{h_i\}$ terms. Under this simplification, the Ising Hamiltonian becomes
\begin{equation}\eqnlabel{IsingHnoh}
	H = - \sum_{i,j,~i < j} J_{ij} s_i s_j.
\end{equation}

What makes the Ising model particularly interesting is that many hard optimisation problems can be shown to be equivalent to it \cite{bian2010ising}.
In fact, all of Karp's 21 NP-complete problems can be mapped to it by assigning appropriate values to the coefficients \cite{lucas2013ising}.
Physical systems that can directly minimise the Ising Hamiltonian, namely Ising machines, thus become very attractive for outperforming conventional algorithms run on CPUs for these problems.

Several schemes have been proposed recently for realising Ising machines in hardware.
One well-known example is from D-Wave Systems \cite{johnson2011quantum,bian2014Ising}.
Their quantum Ising machines use superconducting loops as spins and connect them using Josephson junction devices \cite{harris2010flux}.
As the machines require a temperature below 80mK ($-273.07^\circ$C) to operate \cite{johnson2011quantum}, they all have a large footprint to accommodate the necessary cooling system.
While many question their advantages over simulated annealing run on classical computers \cite{ronnow2014defining}, proponents believe that through a mechanism known as quantum tunnelling, they can offer large speedups on problems with certain energy landscapes \cite{denchev2016tunneling}.

Other proposals use novel non-quantum devices as Ising spins instead, so that the machines can function at room temperature.
Most notable among them is a scheme based on lasers and kilometre long optical fibres \cite{marandi2014network,mcmahon2016ScienceIsing100,inagaki2016ScienceIsing2000}.
The Ising spins are represented using time-multiplexed optical parametric oscillators (OPOs), which are laser pulses travelling on the same fibre.
The coupling between these pulses is implemented digitally by measurement and feedback using an FPGA.
While these machines can potentially be more compact than D-Wave's machines, it is unclear how they can be miniaturised and integrated due to the use of long fibres.
Recent studies have also proposed the use of several novel nanodevices as Ising spins, including MEMS (Micro-Electro-Mechanical Systems) resonators \cite{mahboob2016electromechanical} and nanomagnets from Spintronics \cite{camsari2017pbits}.
Physical realisation of these machines still awaits future development of these emerging device technologies.

Another broad direction is to build Ising model emulators using digital circuits.
A recent implementation \cite{yamaoka2016IsingCMOS} uses CMOS SRAM cells as spins, and couples them using digital logic gates.
The authors point out, however, that ``the efficacy in achieving a global energy minimum is limited'' \cite{yamaoka2016IsingCMOS} due to variability.
The speed-up and accuracy reported by \cite{yamaoka2016IsingCMOS} are instead based on deterministic on-chip computation paired with an external random number generator --- a digital hardware implementation of the simulated annealing algorithm.
More recently, similar digital accelerators have also been tried on FPGAs \cite{yamamoto2017time}.
These implementations are not directly comparable to the other Ising machine implementations discussed above, which attempt to use interesting intrinsic physics to minimise the Ising Hamiltonian for achieving large speedups.

\section{Oscillator-based Ising Machines}\seclabel{main}

In this section, we show that a network of coupled self-sustaining oscillators can function as an Ising machine.
To do so, we first study the response of a single oscillator under injection locking in \secref{Gen-Adler}.
Specifically, we examine the way the oscillator's phase locks to that of the external input.
While regular injection locking typically aligns the oscillator's phase with the input, as illustrated in \figref{fig1} (a) and (b), a variant --- subharmonic injection locking (SHIL) --- can make the oscillator develop multiple stable phase-locked states (\figref{fig1} (c) and (d)).
As we show in \secref{Gen-Adler}, these phenomena can be predicted accurately
using the Gen-Adler equation \cite{BhRoASPDAC2009}.

\begin{figure}[htbp!]
    \vskip1em
    \centering
    {
        \epsfig{file=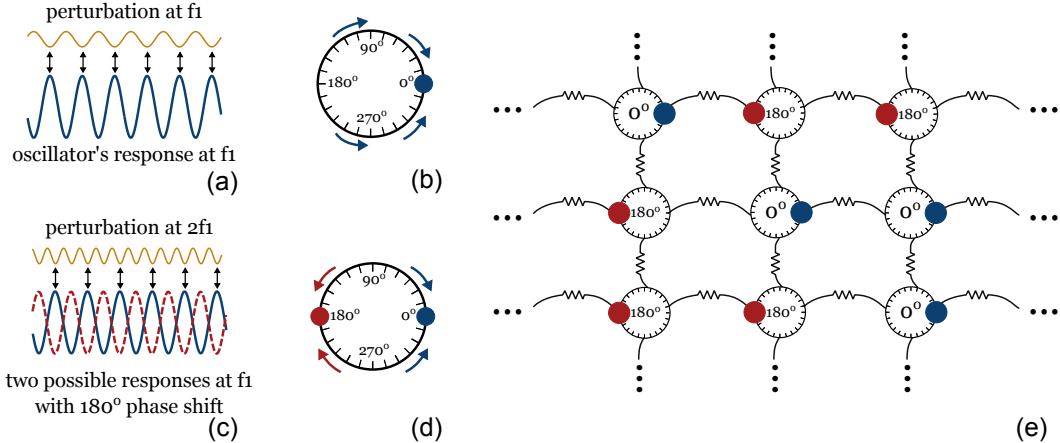, width=\linewidth}
    }
    \caption{Illustration of the basic mechanism of oscillator-based Ising machines:
             (a) oscillator shifts its natural frequency from $f_0$ to $f_1$ under external perturbation;
             (b) oscillator's phase becomes stably locked to the perturbation;
             (c) when the perturbation is at $2f_1$, the oscillator locks to its subharmonic at $f_1$;
             (d) bistable phase locks under subharmonic injection locking;
             (e) coupled subharmonically injection-locked oscillators settle with binary phases
                 representing an optimal spin configuration for an Ising problem.
    \figlabel{fig1}}
    \vskip1em
\end{figure}

The Gen-Adler equation of a single oscillator, when extended to the phase dynamics of coupled oscillator networks, becomes equivalent to a variant of the Kuramoto model.\ignore{, known as the Kuramoto-Sakaguchi model \cite{sakaguchi1986soluble}. JR: not sure of this any more.}
In \secref{Kuramoto}, we show that the model's dynamics are governed by a global Lyapunov function, a scalar ``energy like'' quantity that is naturally minimised by the coupled oscillator network.
Then in \secref{Kuramoto-SHIL}, we introduce SHIL into the system to binarise the phases of oscillators.
As illustrated in \figref{fig1} (e), SHIL induces each oscillator to settle to one of two stable phase-locked states. 
Due to the coupling between them, a network of such binarised oscillators will prefer certain phase configurations over others.
We confirm this intuition in \secref{Kuramoto-SHIL} by deriving a new Lyapunov function that such a system (\ie, with SHIL) minimises.
By examining this function's equivalence to the Ising Hamiltonian, we show that such a coupled oscillator network under SHIL indeed physically implements an Ising machine.
Finally, in \secref{Kuramoto-variability}, we consider the effect of variability on the system's operation. 
We show that a spread in the natural frequencies of the oscillators contributes a linear term in the global Lyapunov function, and does affect Ising machine performance by much if the variability is not extreme.

\subsection{Injection locking in oscillators}\seclabel{Gen-Adler}

When an oscillator with a natural frequency $\omega_0$ is perturbed by a small periodic input at a similar frequency $\omega_1$, its phase response can be predicted well using the Generalised Adler's model (Gen-Adler) \cite{BhRoASPDAC2009}.
Gen-Adler has the following form:
\begin{equation}\eqnlabel{Gen-Adler}
	\frac{d}{dt} \phi(t) = \omega_0-\omega_1 + \omega_0 \cdot c(\phi(t) - \phi_{in}),
\end{equation}
where $\phi(t)$ and $\phi_{in}$ are the phases of the oscillator and the perturbation.\footnote{More rigorous definitions are given in \ref{app:derivation}.}
$c(.)$ is a $2\pi$-periodic function derived based on an intrinsic quantity of the oscillator, known as the Phase Response Curve (PRC) \cite{Winfree67} or the Perturbation Projection Vector (PPV) \cite{DeMeRoTCAS2000}.
% $c(.)$ is not necessarily sinusoidal.
A detailed derivation of Gen-Adler from the low-level differential equations of an oscillator and its PPV is provided in \ref{app:derivation}.

The Gen-Adler equation governs the dynamics of the oscillator's phase under periodic inputs; its equilibrium states can be used to accurately predict the injection-locked states of the oscillator.
The equilibrium Gen-Adler equation can be derived by rearranging \eqnref{Gen-Adler}:
\begin{equation}\eqnlabel{Gen-Adler-equilibrium}
	\frac{\omega_1-\omega_0}{\omega_0} = c(\phi^* - \phi_{in}).
\end{equation}

The Left Hand Side (LHS) of \eqnref{Gen-Adler-equilibrium} is a constant representing the frequency detuning of the oscillator from the input; the Right Hand Side (RHS) is a periodic function of $\phi^*$ whose magnitude depends on both the PPV of the oscillator and the strength of the input \cite{BhRoASPDAC2009}.
By plotting both terms and looking for intersections, one can easily predict whether injection locking will occur, and if it does, what the locked phase of the oscillator will be.
\figref{plot_Adler_annotated} (a) plots a few examples of LHS/RHS, showing their shapes and magnitudes under different conditions.

\begin{figure}[htbp!]
    \centering
    {
        \vspace{0.2em}
        \epsfig{file=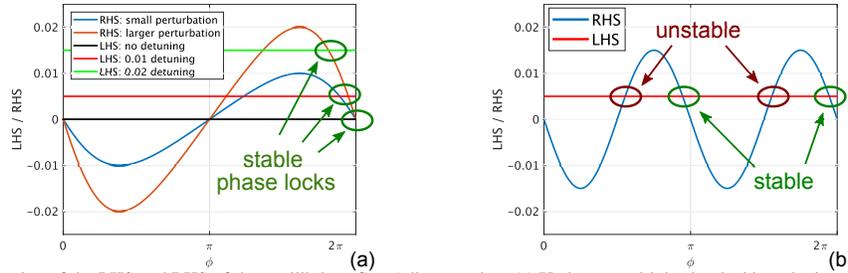, width=0.8\linewidth}
    }
    \caption{Illustration of the LHS and RHS of the equilibrium Gen-Adler equation.
	 (a) Under normal injection locking, the intersection of LHS and RHS
predicts the only solution of $\phi$ under different scenarios.
(b) Perturbation at $2\omega_1$ changes the shape of $c(.)$ in Gen-Adler; the
intersections now predict the locations of two stable phase-locked states.
    \figlabel{plot_Adler_annotated}}
\end{figure}

As mentioned in \secref{intro}, SHIL can occur when the external input is about twice as fast as the oscillator.
When the input is at frequency $2\omega_1$, it can be shown that the corresponding $c(.)$ becomes a $\pi$-periodic function \cite{NeRoDATE2012SHIL,WaRoDAC2015MAPPforPHLOGON}; a typical example is given in \figref{plot_Adler_annotated} (b), where $c(.)$ takes the shape of $-\sin(2\phi)$.
In this case, two of the four LHS-RHS intersections represent stable phase-locked states; it can be shown that they are separated by a phase difference of $180^\circ$ \cite{NeRoDATE2012SHIL}.
Gen-Adler is a powerful technique for predicting and understanding injection locking in oscillators and constitutes an important foundation for the analyses that follow.

\subsection{Global Lyapunov function}\seclabel{Kuramoto}

For an oscillator in a coupled oscillator network, its external perturbations come from the other oscillators that are connected to it.
Its Gen-Adler equation can be written as
\begin{equation}\eqnlabel{Gen-Adler-i}
	\frac{d}{dt} \phi_i(t) = \omega_i - \omega^* + \omega_i \cdot \sum_{j=1,~j \neq i}^n c_{ij}(\phi_i(t) - \phi_j(t)),
\end{equation}
where $\{\phi_i\}$ represents the phases of $n$ oscillators; $\omega_i$ is the frequency of the oscillator whereas $\omega^*$ is the central frequency of the network.
$c_{ij}(.)$ is a $2\pi$-periodic function for the coupling between oscillators $i$ and $j$.

%For coupled oscillators, \eqnref{Gen-Adler-i} is also known as the Kuramoto-Sakaguchi model \cite{sakaguchi1986soluble}.
To simplify exposition, we now assume that the $c_{ij}$ functions are sinusoidal, although in \ref{app:derivation}, we show that this does not have to be the case for the analysis to hold true.\footnote{More generally, $c_{ij}$s can be any $2\pi$-periodic odd functions, which are better suited to practical oscillators.}
We further assume zero spread in the natural frequencies of oscillators, \ie, $\omega_i \equiv \omega^*$, and discuss the effect of frequency variability later in \secref{Kuramoto-variability}.
With these simplifications, \eqnref{Gen-Adler-i} can be written as
\begin{equation}\eqnlabel{Kuramoto}
	\frac{d}{dt} \phi_i(t) = -K \cdot \sum_{j=1,~j \neq i}^n J_{ij} \cdot \sin(\phi_i(t) - \phi_j(t)).
\end{equation}
Here, we are using the coefficients $\{J_{ij}\}$\footnote{In the Ising Hamiltonian \eqnref{IsingH}, $J_{ij}$ is only defined when $i<j$; here we assume that $J_{ij} = J_{ji}$ for all $i,~j$.} from the Ising model \eqnref{IsingH} to set the connectivity of the network, \ie, the coupling strength between oscillators $i$ and $j$ is proportional to $J_{ij}$.
The parameter $K$ modulates the overall coupling strength of the network.

There is a global Lyapunov function associated with \eqnref{Kuramoto} \cite{shinomoto1986phase}:
\begin{equation}\eqnlabel{KuramotoE}
	E(\vec\phi(t)) = -K\cdot \sum_{i,j,~i \neq j} J_{ij} \cdot \cos(\phi_i(t) - \phi_j(t)),
\end{equation}
where $\vec\phi(t) = [\phi_1(t),\cdots,\phi_n(t)]^T$.
Being a global Lyapunov function, it is an objective function the coupled oscillator system always tends to minimise as it evolves over time \cite{lyapunov1992general}.

If we look at the values of this continuous function $E(\vec\phi(t))$ at some discrete points, we notice that it shares some similarities with the Ising Hamiltonian.
At points where every $\phi_i$ is equal to either $0$ or $\pi$,\footnote{More generally, we can use $\{2k\pi~|~k\in \mathbf{Z}\}$ and $\{2k\pi+\pi~|~k\in \mathbf{Z}\}$ to represent the two states for each oscillator's phase.} if we map $\phi_i=0$ to $s_i=+1$ and $\phi_i=\pi$ to $s_i=-1$, we have
\begin{equation}
	E(\vec\phi(t)) = - K\cdot \sum_{i,j,~i \neq j} J_{ij} \cdot \cos(\phi_i(t) - \phi_j(t))
	= - K\cdot \sum_{i,j,~i \neq j} J_{ij} s_i s_j
	= - 2K\cdot \sum_{i,j,~i < j} J_{ij} s_i s_j.
\end{equation}

If we choose $K = 1/2$, the global Lyapunov function in \eqnref{KuramotoE} exactly matches the Ising Hamiltonian in \eqnref{IsingHnoh} at these discrete points.
But this does not mean that coupled oscillators are naturally minimising the Ising Hamiltonian, as there is no guarantee at all that the phases $\{\phi_i(t)\}$ are settling to these discrete points.
In fact, networks with more than two oscillators almost always synchronise with analog phases, \ie, $\{\phi_i(t)\}$ commonly settle to continuous values spread out in the phase domain as opposed to converging towards $0$ and $\pi$.
As an example, \figref{Kuramoto_with_without_SYNC} (a) shows the phase responses of 20 oscillators connected in a random graph.
As phases do not settle to the discrete points discussed above, the Lyapunov function they minimise becomes irrelevant to the Ising Hamiltonian, rendering the system ineffective for solving Ising problems.
While one may think that the analog phases can still serve as solutions when rounded to the nearest discrete points, experiments in \secref{largerMAXCUT} show that the quality of these solutions is very poor compared with our scheme of Ising machines proposed in this paper.

\begin{figure}[htbp!]
    \centering
    {
        \epsfig{file=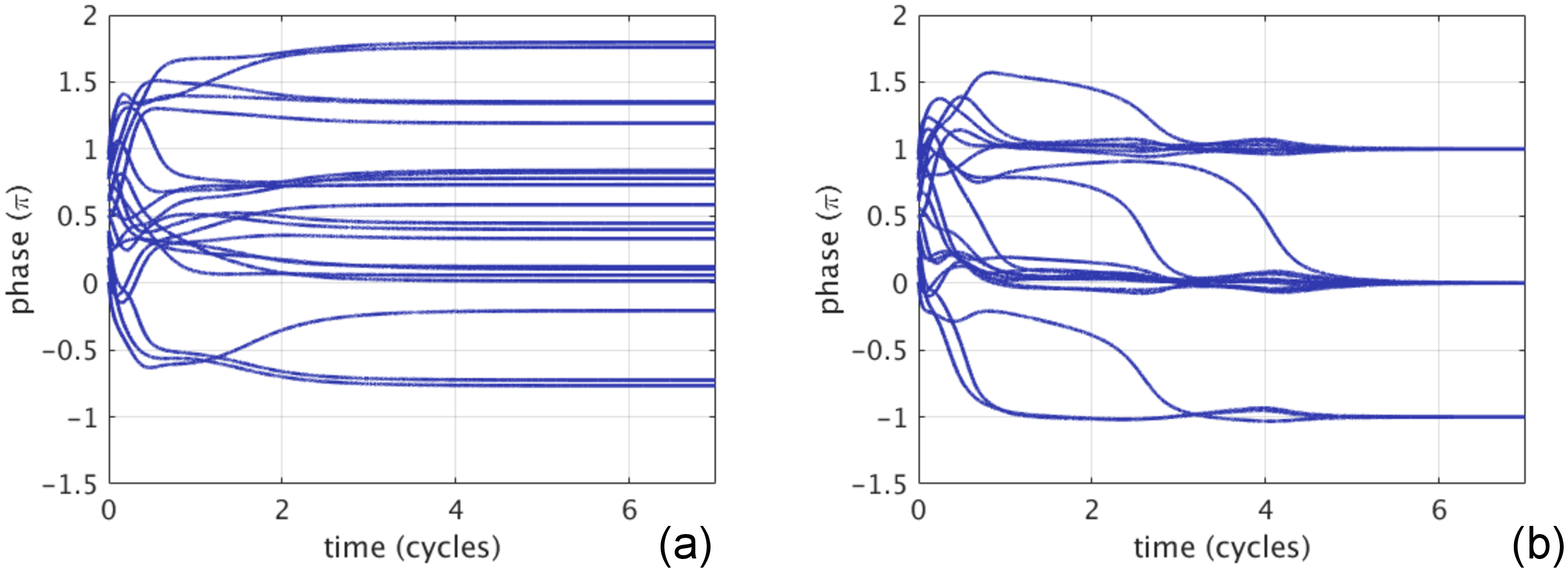, width=0.8\linewidth}
    }
    \caption{Phases of 20 oscillators with random $\{J_{ij}\}$ generated by \texttt{rudy -rnd\_graph 20 50 10001}: \\ (a) without SYNC; (b) with $K_s = 1$. % encoding a random MAX-CUT problem generated by ...
    \figlabel{Kuramoto_with_without_SYNC}}
\end{figure}

\subsection{Network of coupled oscillators under SHIL and its global Lyapunov function }\seclabel{Kuramoto-SHIL}

In our scheme, a common SYNC signal at $2\omega^*$ is injected to every
oscillator in the network. 
Through the mechanism of SHIL, the oscillator phases are binarised.
The example shown in \figref{Kuramoto_with_without_SYNC} (b) confirms that this is indeed the case: under SHIL, the phases of 20 oscillators connected in the same random graph now settle very close to discrete points.
To write the model for such a system, we recall from \secref{Gen-Adler} that a
$2\omega^*$ perturbation introduces a $\pi$-periodic coupling term (\eg,
$\sin(2\phi)$) in the phase dynamics.
Therefore, here we can directly write the model as follows and show its
derivation in \ref{app:derivation}.
\begin{equation}\eqnlabel{KuramotoSHIL}
	\frac{d}{dt} \phi_i(t) = - K\cdot \sum_{j=1,~j\neq i}^n J_{ij} \cdot \sin(\phi_i(t) - \phi_j(t))
						- K_s \cdot \sin(2\phi_i(t)),
\end{equation}
where $K_s$ represents the strength of coupling from SYNC.

Remarkably, there is a global Lyapunov function for this new type of coupled oscillator system.
It can be written as
\begin{equation}\eqnlabel{ESHIL1}
	E(\vec\phi(t)) = - K\cdot \sum_{i,j,~i\neq j} J_{ij} \cdot \cos(\phi_i(t) - \phi_j(t))
	  - K_s \cdot \sum_{i=1}^n \cos\left(2\phi_i(t)\right).
\end{equation}

Now, we show that $E$ in \eqnref{ESHIL1} is indeed a global Lyapunov function.
To do so, we first differentiate $E$ with respect to $\vec \phi$.
We observe that the first component of $E$ is the sum of $(n^2-n)$ number of $\cos()$ terms.
Among them, for any given index $k$, variable $\phi_k$ appears a total of
$2\cdot(n-1)$ times.
It appears $(n-1)$ times as the subtrahend inside $\cos()$: these $(n-1)$ terms
are $J_{kl} \cdot \cos(\phi_k(t) - \phi_l(t))$, where $l=1,\cdots,n$ and $l\neq k$.
For the other $(n-1)$ times, it appears as the minuend inside $\cos()$: in
$J_{lk} \cdot \cos(\phi_l(t) - \phi_k(t))$, where $l=1,\cdots,n$, $l\neq k$.
So when we differentiate $E$ with respect to $\phi_k$, we have
\begin{align}
	\frac{\partial E(\vec\phi(t))}{\partial \phi_k(t)} =& -K\cdot \sum_{l=1,~l\neq k}^n J_{kl}
					 \frac{\partial}{\partial\phi_k(t)} \left[\cos(\phi_k(t) - \phi_l(t))\right]
	                        -K\cdot \sum_{l=1,~l\neq k}^n J_{lk}
					 \frac{\partial}{\partial\phi_k(t)} \left[\cos(\phi_l(t) - \phi_k(t))\right]  \nonumber \\
                     & - K_s\cdot \frac{\partial}{\partial \phi_k(t)}\cos(2\phi_k(t)) \\
    =& K\cdot \sum_{l=1,~l\neq k}^n J_{kl}
					 \sin(\phi_k(t) - \phi_l(t))
	                        - K\cdot \sum_{l=1,~l\neq k}^n J_{lk}
					 \sin(\phi_l(t) - \phi_k(t)) + K_s\cdot 2 \cdot \sin(2\phi_k(t)) \\
    =& K\cdot \sum_{l=1,~l\neq k}^n J_{kl} \cdot 2 \cdot \sin(\phi_k(t) - \phi_l(t))
     + K_s\cdot 2 \cdot \sin(2\phi_k(t)) \\
    =& -2 \cdot \frac{d\phi_k(t)}{dt}.
\end{align}

Therefore, 
\begin{align}
	\frac{\partial E(\vec\phi(t))}{\partial t} &= \sum_{k=1}^n \left[ \frac{\partial E(\vec\phi(t))}{\partial \phi_k(t)} \cdot
                     \frac{d\phi_k(t)}{dt} \right] \\
	              &= - 2 \cdot \sum_{k=1}^n \left(\frac{d\phi_k(t)}{dt}\right)^2 \leq 0.
\end{align}

Thus, we have proved that \eqnref{ESHIL1} is indeed a global Lyapunov function the coupled oscillators under SHIL naturally minimise over time.
A similar but more detailed proof for the general case where we do not assume
sinusoidal coupling functions is given in \ref{app:proof}.

At the discrete points (phase values of 0/$\pi$), because $\cos(2\phi_i) \equiv 1$, \eqnref{ESHIL1} reduces to
\begin{equation}\eqnlabel{ESHIL2}
	E(\vec\phi(t)) \approx - K\cdot \sum_{i,j,~i\neq j} J_{ij} \cdot \cos(\phi_i(t) - \phi_j(t))
	  - n\cdot K_s,
\end{equation}
where $n\cdot K_s$ is a constant.
By choosing $K=1/2$, we can then make \eqnref{ESHIL2} equivalent to the Ising Hamiltonian in \eqnref{IsingHnoh} with a constant offset.

Note that the introduction of SYNC does not change the relative $E$ levels between the discrete points, but modifies them by the \textit{same} amount.
However, with SYNC, all phases can be forced to eventually take values near either $0$ or $\pi$ --- the system now tries to reach a binary state that minimises the Ising Hamiltonian, thus functioning as an Ising machine.
We emphasise that this is \textit{not} equivalent to running the system without SHIL and then rounding the analog phase solutions to discrete values as a post-processing step. 
Instead, the introduction of SHIL modifies the energy landscape of $E$, changes the dynamics of the coupled oscillator system, and as we show in \secref{examples}, results in greatly improved minimisation of the Ising Hamiltonian. 

It is worth noting, also, that the Lyapunov function in \eqnref{ESHIL1} will, in general, have many local minima and there is no guarantee the oscillator-based Ising machine will settle at or near any global optimal state.
However, as we show in \ref{app:noise}, when judicious amounts of noise are introduced via a noise level parameter $K_n$, it becomes more likely to settle to lower minima.
\ignore{An appropriate noise level $K_n$ can improve the performance of the Ising machine.}
Indeed, the several parameters in the Ising machine --- $K$, $K_s$ and $K_n$ --- all play an important role in its operation and should be given suitable values.
Furthermore, $K$, $K_s$, $K_n$ can also be time varying, creating various ``annealing schedules''.
As we show in \secref{examples}, this feature gives us considerable flexibility in operating oscillator-based Ising machines for good performance.

\subsection{Coupled oscillator networks with frequency variations}\seclabel{Kuramoto-variability}

A major obstacle to practical implementation of large-scale Ising machines is variability.
While few analyses exist for assessing the effects of variability for previous Ising machine technologies (\secref{IsingIntro}), the effect of variability on our oscillator-based Ising machine scheme is easy to analyze, predicting that performance degrades gracefully.

One very attractive feature of oscillators is that variability, regardless of the nature and number of elemental physical sources, eventually manifests itself essentially in only one parameter, namely the oscillator's natural frequency.
As a result, the effect of variability in an oscillator network is that there is a spread in the natural frequencies of the oscillators.
Taking this into consideration, our model can be revised as
\be{KuramotoVar}
\frac{d}{d t} \phi_i(t) = ~\omega_i-\omega^* - \omega_i \cdot K \cdot
        \sum_{j=1,~j\neq i}^n J_{ij} \cdot \sin(\phi_i(t) -  \phi_j(t))
        - \omega_i \cdot K_s \cdot \sin(2\phi_i(t)).
\ee

As it turns out, there is also a global Lyapunov function associated with this system.
\begin{equation}\eqnlabel{EVar}
	E(\vec\phi(t)) = - K\cdot \sum_{i,j,~i\neq j} J_{ij} \cdot \cos(\phi_i(t) - \phi_j(t))
	  - K_s \cdot \sum_{i=1}^n \cos\left(2\phi_i(t)\right)
      - 2\sum_{i=1}^n \frac{\omega_i - \omega^*}{\omega_i}\phi_i.
\end{equation}

This can be proven as follows.
\begin{align}
	\frac{\partial E(\vec\phi(t))}{\partial \phi_k(t)}
    =& K\cdot \sum_{l=1,~l\neq k}^n J_{kl} \cdot 2 \cdot \sin(\phi_k(t) - \phi_l(t))
     + K_s\cdot 2 \cdot \sin(2\phi_k(t)) 
      - 2\frac{\omega_k - \omega^*}{\omega_k} \\
    =& -\frac{2}{\omega_k} \cdot \frac{d\phi_k(t)}{dt}.
\end{align}
Therefore, 
\begin{equation}
	\frac{dE(\vec\phi(t))}{dt} = - \sum_{k=1}^n \frac{2}{\omega_k}\left(\frac{d\phi_k(t)}{dt}\right)^2 \leq 0.
\end{equation}

Note that \eqnref{EVar} differs from \eqnref{ESHIL1} only by a weighted sum of $\phi_i$ --- it represents essentially the same energy landscape but tilted linearly with the optimisation variables.
While it can still change the locations and values of the solutions, its effects are easy to analyse given a specific combinatorial optimisation problem.
Also, as the coupling coefficient $K$ gets larger, the effect of variability can be reduced.
Small amounts of variability merely perturb the locations of minima a little, \ie, the overall performance of the Ising machine remains essentially unaffected.
Very large amounts of variability can, of course, eliminate minima that would exist if there were no variability.
However, another great advantage of using oscillators is that even in the presence of large variability, the frequency oscillators can be calibrated (\eg, using a voltage-controlled oscillator (VCO) scheme) prior to each run of the machine. 
As a result, the spread in frequencies can be essentially eliminated in a practical and easy-to-implement way.

\section{Examples}\seclabel{examples}

In this section, we demonstrate the feasibility and efficacy of our oscillator-based Ising machine scheme by applying it to several MAX-CUT examples and a graph-colouring problem.

\subsection{Small MAX-CUT Problems}\seclabel{smallMAXCUT}

Given an undirected graph, the MAX-CUT problem \cite{myklebust2015SA,festa2002randomized} asks us to find a subset of vertices such that the total weights of the cut set between this subset and the remaining vertices are maximised.
As an example, \figref{maxcut-cubic-graph} shows a size-8 cubic graph, where each vertex is connected to three others --- neighbours on both sides and the opposing vertex.
As shown in \figref{maxcut-cubic-graph}, dividing the 8 vertices randomly yields a cut size of 5; grouping even and odd vertices, which one may think is the best strategy, results in a cut size of 8; the maximum cut is actually 10, with one of the solutions shown in the illustration.
Changing the edge weights to non-unit values can change the maximum cut and also make the solution look less regular, often making the problem more difficult to solve.
While the problem may not seem challenging at size 8, it quickly becomes intractable as the size of the graph grows.
In fact, MAX-CUT is one of Karp's 21 NP-complete problems \cite{karp1972np}.

\begin{figure}[htbp!]
    \centering
    {
        \epsfig{file=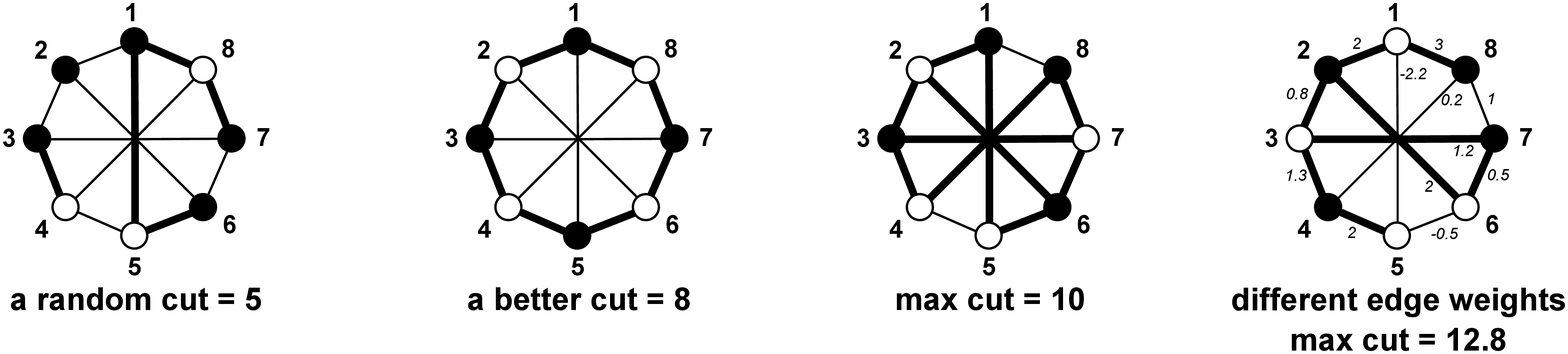, width=\linewidth}
    }
	\caption{Illustration of different cut sizes in a 8-vertex cubic graph with unit
             edge weights, and another one with random weights (rightmost). 
    \figlabel{maxcut-cubic-graph}}
\end{figure}

The MAX-CUT problem has a direct mapping to the Ising model \cite{karp1972np}, by choosing $J_{ij}$ to be the opposite of the weight of the edge between vertices $i$ and $j$, \ie, $J_{ij} = -w_{ij}$.
To explain this mapping scheme, we can divide the vertices into two sets --- $V_1$ and $V_2$.
Accordingly, all the edges in the graph are separated into three groups --- those that connect vertices within $V_1$, those within $V_2$, and the cut set containing edges across $V_1$ and $V_2$.
The sums of the weights in these three sets are denoted by $S_1$, $S_2$ and $S_{cut}$.
Together, they constitute the total edge weights of the graph, which is also the negation of the sum of all the $J_{ij}$s:
\begin{equation}
	 S_1 + S_2 + S_{cut} = \sum_{i,j,~i<j} w_{ij}  = -\sum_{i,j,~i<j} J_{ij}.
\end{equation}

We then map this division of vertices to the values of Ising spins, assigning $+1$ to a spin $i$ if vertex $v_i\in V_1$, and $-1$ if the vertex is in $V_2$.  
The Ising Hamiltonian in \eqnref{IsingHnoh} can then be calculated as
\begin{align}
	H =& \sum_{i,j,~i<j} J_{ij} s_i s_j 
    \nonumber \\
      =& \sum_{i<j,~v_i,v_j\in V_1} J_{ij} (+1)(+1) +
	\sum_{i<j,~v_i,v_j\in V_2} J_{ij} (-1)(-1) + \sum_{i<j,~v_i\in V_1,v_j\in V_2} J_{ij} (+1)(-1) 
    \nonumber \\
	=& \sum_{i<j,~v_i,v_j\in V_1} J_{ij} + \sum_{i<j,~v_i,v_j\in V_2} J_{ij} - \sum_{i<j,~v_i\in V_1,j\in V_2} J_{ij} 
    \nonumber \\
    =& -(S_1 + S_2 - S_{cut}) = \sum_{i,j,~i<j} J_{ij} - 2 \cdot S_{cut}.
\end{align}

Therefore, when the Ising Hamiltonian is minimised, the cut size is maximised.

To show that an oscillator-based Ising machine can indeed be used to solve MAX-CUT problems, we simulated the Kuramoto model in \eqnref{KuramotoSHIL} while making the $J_{ij}$s represent the unit-weight cubic graph in \figref{maxcut-cubic-graph}.
The magnitude of SYNC is fixed at $K_s=3$, while we ramp up the coupling strength $K$ from 0 to 5.
Results from the deterministic model ($K_n=0$) and the stochastic model ($K_n=0.1$) are shown in \figref{MAXCUT8} and \figref{MAXCUT8-noise} respectively.
In the simulations, oscillators started with random phases between 0 and $\pi$; after a while, they all settled to one of the two phase-locked states separated by $\pi$.
These two groups of oscillators represent the two subsets of vertices in the solution.
The results for the 8 spins shown in \figref{MAXCUT8} and \figref{MAXCUT8-noise} are $\{+1,-1,+1,-1,-1,+1,-1,+1\}$ and $\{-1,+1,+1,-1,+1,-1,-1,+1\}$ respectively; both are global optimal solutions.

\begin{figure}[htbp]
	\begin{minipage}{0.45\linewidth}
      \epsfig{file=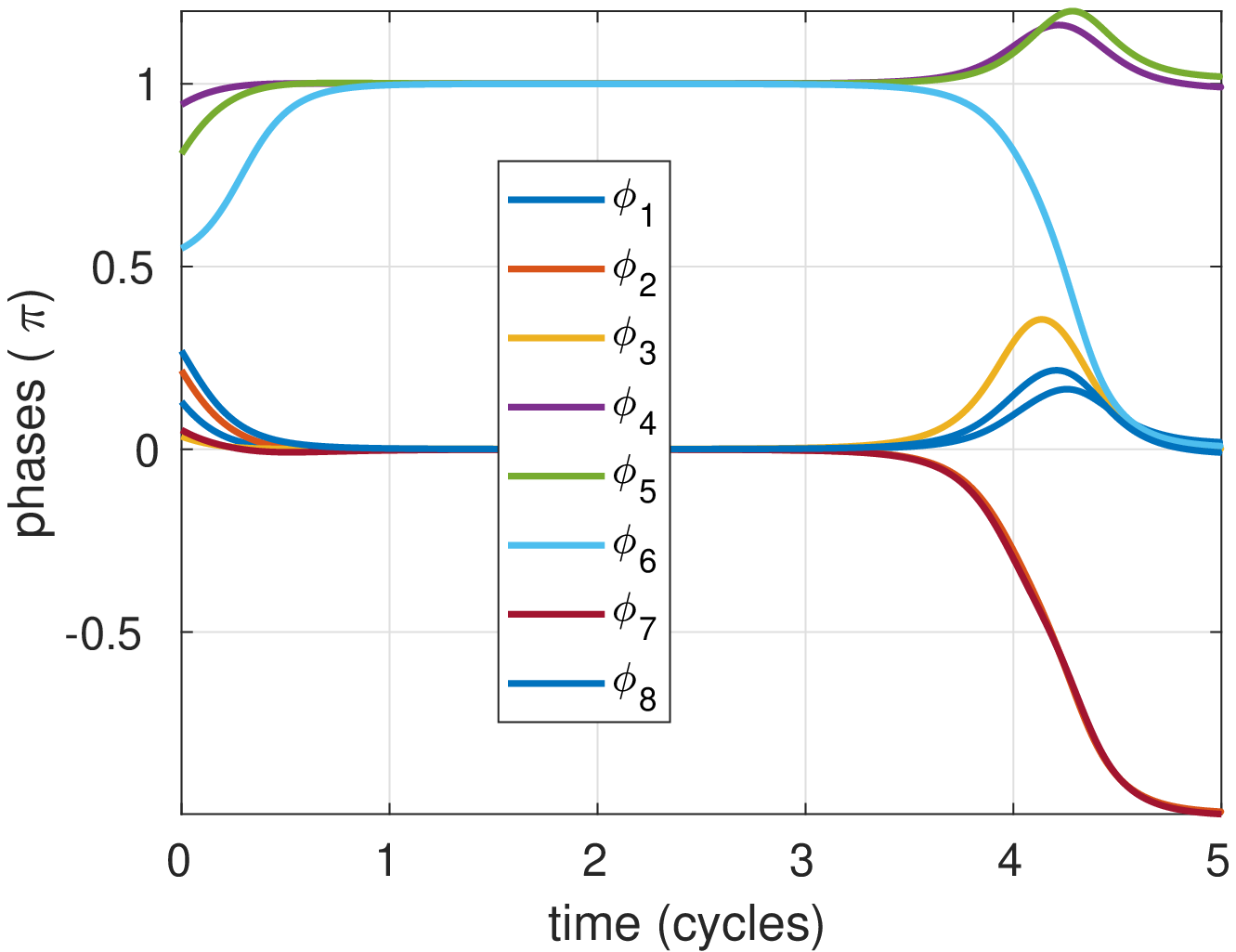,width=\linewidth}
	  \caption{Phases of oscillators solving a size-8 MAX-CUT problem without noise.}\figlabel{MAXCUT8}
	\end{minipage}
    \hfill
	\begin{minipage}{0.45\linewidth}
      \epsfig{file=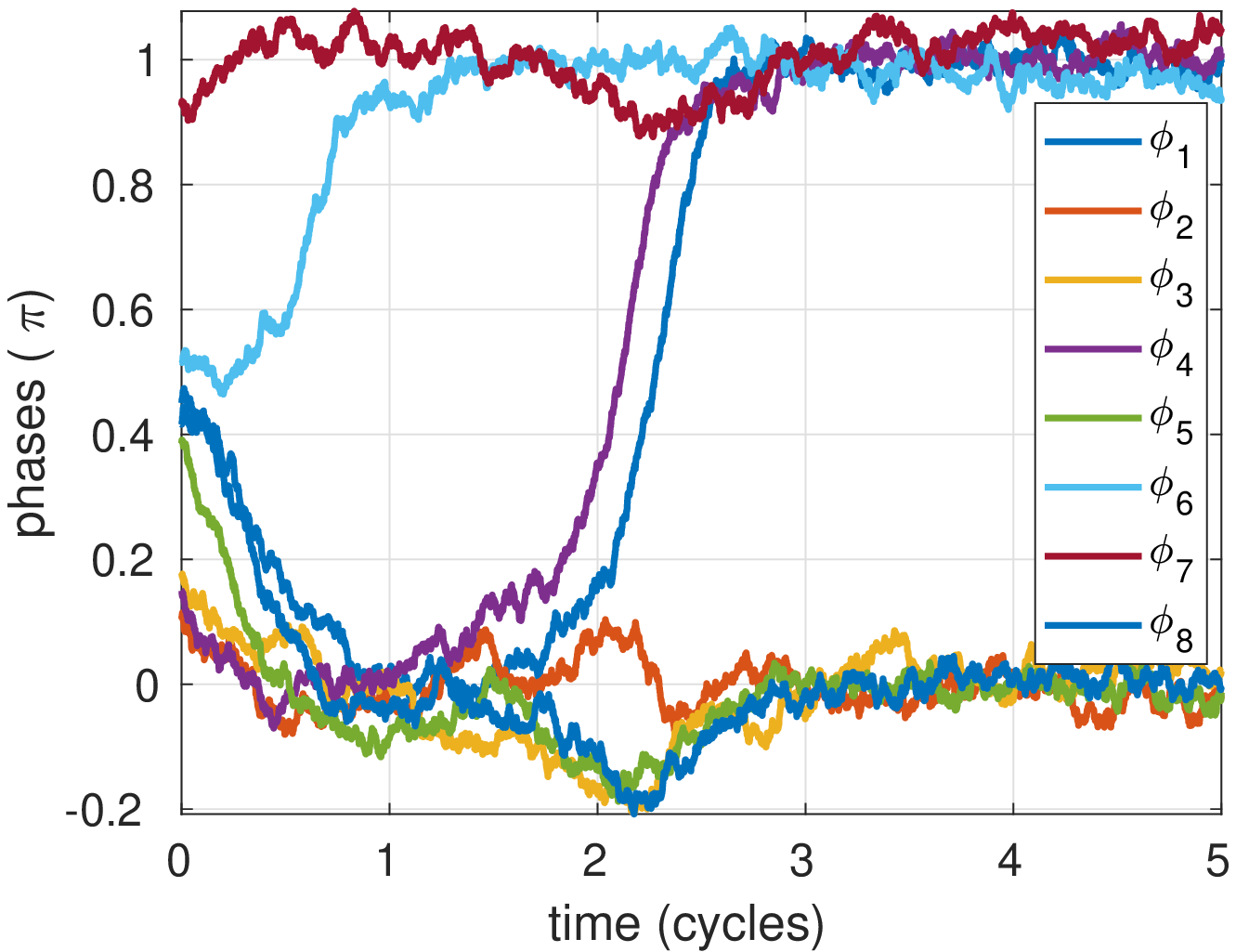,width=\linewidth}
	  \caption{Phases of oscillators solving a size-8 MAX-CUT problem with noise.}\figlabel{MAXCUT8-noise}
	\end{minipage}
\end{figure}

A minimal code for reproducing these results is show in \ref{app:code} %. 
Note that these are simulations on stochastic differential equations with random initial conditions.
Every run will return different waveforms; there is no guarantee that the global optimum will be reached on every run.

\begin{figure}[htbp!]
    \centering
    {
        \epsfig{file=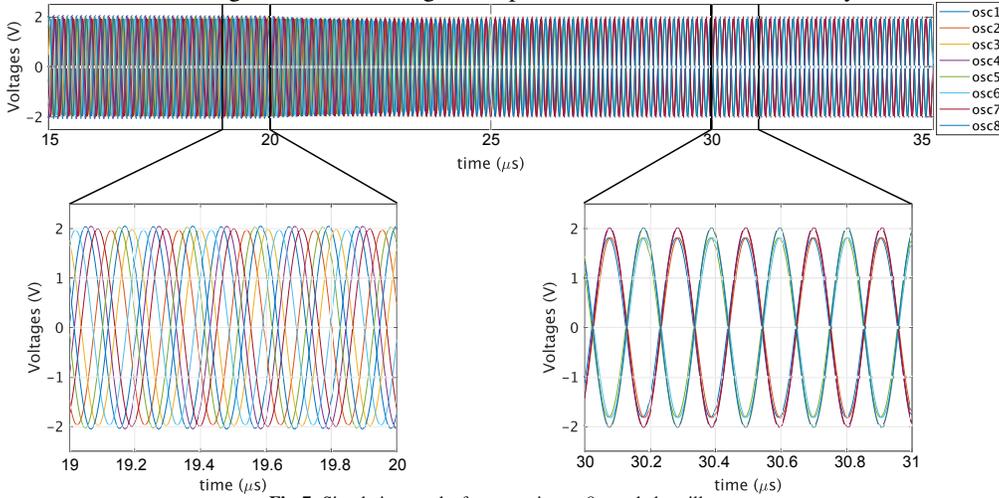, width=0.95\linewidth}
    }
    \caption{Simulation results from ngspice on 8 coupled oscillators.
    \figlabel{MAXCUT_LC}}
\end{figure}
We have also directly simulated coupled oscillators at the SPICE level to confirm the results obtained on phase macromodels.
Such simulations are at a lower lever than phase macromodels and are less efficient.
But they are closer to physical reality and are useful for circuit design.  
In the simulations, 8 cross-coupled LC oscillators are tuned to a frequency of 5MHz.
They are coupled through resistors, with conductances proportional to the coupling coefficients; in this case, we use $J_{ij} \cdot 1/100k\Omega$.
% TODO: figure needed
Results from transient simulation using ngspice-28 are shown in \figref{MAXCUT_LC}.
The 8 oscillators' phases settle into the two groups \{1,4,6,7\} and \{2,3,5,8\}, representing one of the optimal solutions for the MAX-CUT problem.
They synchronise within 20$\mu$s after oscillation starts, which is about 100 cycles.
We have tried this computational experiment with different random initial conditions; like phase-macromodels, the SPICE-level simulations of these coupled oscillators reliably return optimal solutions for this size-8 MAX-CUT problem.

\begin{figure}[htbp!]
    \centering
    {
        \epsfig{file=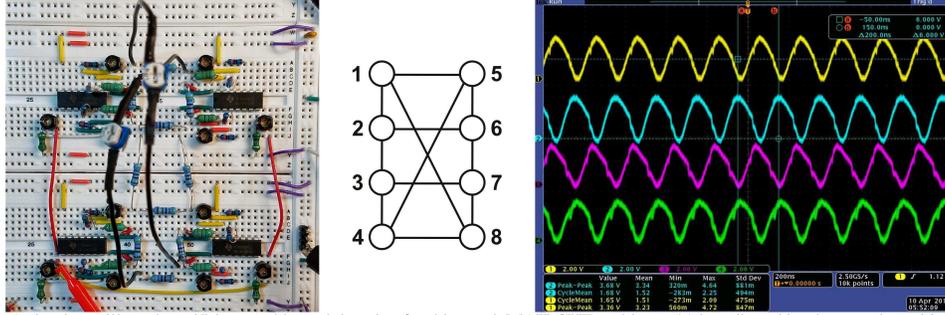, width=0.9\linewidth}
    }
    \caption{A simple oscillator-based Ising machine solving size-8 cubic graph MAX-CUT
problems: (a) breadboard implementation with 8 CMOS LC oscillators;
(b) illustration of the connections; 
(c) oscilloscope measurements showing waveforms of oscillator 1$\sim$4.
    \figlabel{size-8}
    }
\end{figure}
We have also implemented these 8 coupled LC oscillators on a breadboard; a photo of it is shown in \figref{size-8}. 
The inductance of the LC oscillators is provided by fixed inductors of size 33$\mu$H.
The capacitance is provided by trimmer capacitors with a maximum value of 50pF; we have tuned them to around 30pF such that the natural frequencies of all oscillators are about 5MHz.
The nonlinearity for sustaining the LC oscillation is implemented by cross-coupled CMOS inverters on TI SN74HC04N chips.
SYNC is supplied through the GND pins of these chips.
The results have been observed using two four-channel oscilloscopes; a screenshot of one of them is shown in \figref{size-8}.
Through experiments with various sets of edge weights, we have validated that this is indeed a proof-of-concept hardware implementation of oscillator-based Ising machines for size-8 cubic-graph MAX-CUT problems.

\begin{figure}[htbp!]
    \centering
    {
        \epsfig{file=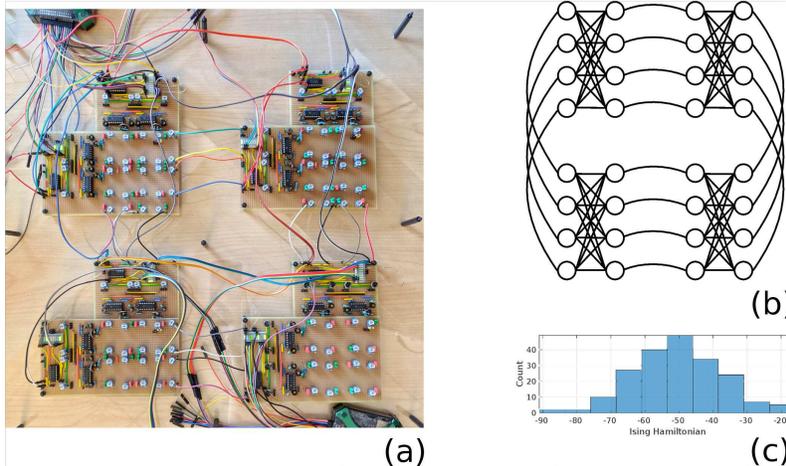, width=0.75\linewidth}
    }
    \caption{
        A size-32 oscillator-based Ising machine:
        (a) photo of the implementation on perfboards;
        (b) illustration of the connectivity ;
        (c) a typical histogram of the energy values achieved in 200 runs on a random
        size-32 Ising problem; the lowest energy level is -88 and is achieved once in
        this case.
        \figlabel{Chimera32}
    }
\end{figure}

Using the same type of oscillators, we have built hardware Ising machines of larger sizes.
\figref{Chimera32} shows a size-32 example implementing a type of connectivity known as the Chimera graph, much like the quantum Ising machines manufactured by D-Wave Systems.
In this graph, oscillators are organised into groups of 8, with denser connections within the groups and sparse ones in between.
The hardware is on perfboards, with components soldered on the boards so that the setup is more permanent than those on breadboards.
Connections are implemented using rotary potentiometers.
Next to each potentiometer we have designed male pin connectors soldered on the board such that the polarity of each connection can be controlled by shorting different pins using female jumper caps.
When encoding Ising problems, we have also colour-coded the jumper caps to make debugging easier, as can be seen in the photo as red and green dots next to the four arrays of white round potentiometers.
To read the phases of the oscillators, instead of using multichannel oscilloscopes, we have soldered TI SN74HC86N Exclusive-OR (XOR) gate chips on board. 
The XOR operation of an oscillator's response and a reference signal converts the oscillating waveform into a high or low voltage level, indicating if the oscillator's phase is aligned with or opposite to the reference phase.
The voltage level can then be picked up by a multichannel logic analyzer.
The entire setup is powered by two Digilent Analog Discovery 2 devices, which are portable USB devices that integrate power supplies, logic analyzers and function generators.
We have tried random Ising problems by programming each connection with a random polarity using the jumper caps.
A typical histogram of the Ising Hamiltonians achieved is shown in \figref{Chimera32} (c).
Note that because $J_{ij}$s have random polarities, a random solution would have an average energy level of zero.
In comparison, the results measured from the hardware are always below 0, and sometimes achieve the global minimum.
While such a hand-soldered system is nontrivial to assemble and operate, and its size of 32 cannot be characterised as large scale, it is a useful proof of concept for implementing oscillator-based Ising machines using standard CMOS technologies, and serves as a very solid basis for our future plans to scale the implementations with custom PCBs and custom ICs.
\subsection{MAX-CUT Benchmark Problems}\seclabel{largerMAXCUT}

In this section, we demonstrate the efficacy of oscillator-based Ising machines for solving larger-scale MAX-CUT problems.
Specifically, we have run simulations on all the problems in a widely used set of MAX-CUT benchmarks known as the G-set \cite{helmberg2000spectral}.\footnote{The G-set problems are available for download as set1 at http://www.optsicom.es/maxcut.}
Problem sizes range from 800 to 3000.\footnote{G1$\sim$21 are of size 800; G22$\sim$42 are of size 2000; G43$\sim$47, G51$\sim$54 are of size 1000; G48$\sim$50 are of size 3000.}
In the experiments, we operated the Ising machine for all the problems with a single annealing schedule, \ie, we did not tune our Ising machine parameters individually for different problems.
Each problem was simulated with 200 random instances.
In \tabref{results}, we list the results and runtime alongside those from several heuristic algorithms developed for MAX-CUT --- Scatter Search (SS) \cite{marti2009SS}, CirCut \cite{burer2002circut}, and Variable Neighbourhood Search with Path Relinking (VNSPR) \cite{festa2002randomized}.\footnote{Their results and runtime are available for download at http://www.optsicom.es/maxcut in the ``Computational Experiences'' section.}
We also list the performances of simulated annealing from a recent study \cite{myklebust2015SA}, the only one we were able to find that contains results for all the G-set problems.

\begin{table}[ht!]
    \begin{center}
% {\footnotesize
{\scriptsize
        \begin{tabular}{|c|cc|cc|cc|cc|ccccc|}
        \hline
{\bf Benchmark} & {\bf SS} & {\bf Time} & {\bf CirCut} & {\bf Time} & {\bf VNSPR} & {\bf Time} & {\bf SA} & {\bf Time} & {\bf OIM} & {\bf Time} & {\bf ~~~~~~    } & {\bf $n_{max}~$} & {\bf $~n_{0.999}$} \\
        \hline
G1 & {\bf 11624} & 139 & {\bf 11624} & 352 & 11621 & 22732 & 11621 & 295 & {\bf 11624} & 52.6 &  & 43 & 123 \\ \hline
G2 & {\bf 11620} & 167 & 11617 & 283 & 11615 & 22719 & 11612 & 327 & {\bf 11620} & 52.7 &  & 1 & 87 \\ \hline
G3 & {\bf 11622} & 180 & {\bf 11622} & 330 & {\bf 11622} & 23890 & 11618 & 295 & {\bf 11622} & 52.4 &  & 10 & 117 \\ \hline
G4 & {\bf 11646} & 194 & 11641 & 524 & 11600 & 24050 & 11644 & 294 & {\bf 11646} & 52.7 &  & 20 & 133 \\ \hline
G5 & {\bf 11631} & 205 & 11627 & 1128 & 11598 & 23134 & 11628 & 300 & {\bf 11631} & 52.6 &  & 3 & 121 \\ \hline
G6 & 2165 & 176 & {\bf 2178} & 947 & 2102 & 18215 & {\bf 2178} & 247 & {\bf 2178} & 52.8 &  & 4 & 46 \\ \hline
G7 & 1982 & 176 & 2003 & 867 & 1906 & 17716 & {\bf 2006} & 205 & 2000 & 52.9 &  & 17 & 21 \\ \hline
G8 & 1986 & 195 & 2003 & 931 & 1908 & 19334 & {\bf 2005} & 206 & 2004 & 52.8 &  & 2 & 26 \\ \hline
G9 & 2040 & 158 & 2048 & 943 & 1998 & 15225 & {\bf 2054} & 206 & {\bf 2054} & 52.6 &  & 2 & 2 \\ \hline
G10 & 1993 & 210 & 1994 & 881 & 1910 & 16269 & 1999 & 205 & {\bf 2000} & 52.9 &  & 21 & 58 \\ \hline
G11 & 562 & 172 & 560 & 74 & {\bf 564} & 10084 & {\bf 564} & 189 & {\bf 564} & 6.7 &  & 6 & 6 \\ \hline
G12 & 552 & 242 & 552 & 58 & {\bf 556} & 10852 & 554 & 189 & {\bf 556} & 6.3 &  & 25 & 25 \\ \hline
G13 & 578 & 228 & 574 & 62 & 580 & 10749 & 580 & 195 & {\bf 582} & 6.4 &  & 3 & 3 \\ \hline
G14 & 3060 & 187 & 3058 & 128 & 3055 & 16734 & {\bf 3063} & 252 & 3061 & 14.6 &  & 27 & 91 \\ \hline
G15 & {\bf 3049} & 143 & {\bf 3049} & 155 & 3043 & 17184 & {\bf 3049} & 220 & {\bf 3049} & 16.1 &  & 41 & 145 \\ \hline
G16 & 3045 & 162 & 3045 & 142 & 3043 & 16562 & 3050 & 219 & {\bf 3052} & 14.5 &  & 8 & 53 \\ \hline
G17 & 3043 & 313 & 3037 & 366 & 3030 & 18555 & 3045 & 219 & {\bf 3046} & 14.6 &  & 5 & 52 \\ \hline
G18 & 988 & 174 & 978 & 497 & 916 & 12578 & {\bf 990} & 235 & {\bf 990} & 14.7 &  & 3 & 3 \\ \hline
G19 & 903 & 128 & 888 & 507 & 836 & 14546 & 904 & 196 & {\bf 906} & 14.5 &  & 13 & 13 \\ \hline
G20 & {\bf 941} & 191 & {\bf 941} & 503 & 900 & 13326 & {\bf 941} & 195 & {\bf 941} & 14.7 &  & 160 & 160 \\ \hline
G21 & 930 & 233 & {\bf 931} & 524 & 902 & 12885 & 927 & 195 & {\bf 931} & 14.6 &  & 10 & 10 \\ \hline
G22 & 13346 & 1336 & 13346 & 493 & 13295 & 197654 & 13158 & 295 & {\bf 13356} & 58.7 &  & 3 & 93 \\ \hline
G23 & 13317 & 1022 & 13317 & 457 & 13290 & 193707 & 13116 & 288 & {\bf 13333} & 58.6 &  & 8 & 54 \\ \hline
G24 & 13303 & 1191 & 13314 & 521 & 13276 & 195749 & 13125 & 289 & {\bf 13329} & 59.0 &  & 6 & 23 \\ \hline
G25 & 13320 & 1299 & {\bf 13326} & 1600 & 12298 & 212563 & 13119 & 316 & {\bf 13326} & 58.7 &  & 6 & 45 \\ \hline
G26 & 13294 & 1415 & {\bf 13314} & 1569 & 12290 & 228969 & 13098 & 289 & 13313 & 58.9 &  & 4 & 81 \\ \hline
G27 & 3318 & 1438 & 3306 & 1456 & 3296 & 35652 & {\bf 3341} & 214 & 3323 & 59.0 &  & 18 & 24 \\ \hline
G28 & 3285 & 1314 & 3260 & 1543 & 3220 & 38655 & {\bf 3298} & 252 & 3285 & 61.2 &  & 1 & 5 \\ \hline
G29 & 3389 & 1266 & 3376 & 1512 & 3303 & 33695 & 3394 & 214 & {\bf 3396} & 58.9 &  & 2 & 8 \\ \hline
G30 & 3403 & 1196 & 3385 & 1463 & 3320 & 34458 & {\bf 3412} & 215 & 3402 & 59.0 &  & 12 & 16 \\ \hline
G31 & 3288 & 1336 & 3285 & 1448 & 3202 & 36658 & {\bf 3309} & 214 & 3296 & 59.1 &  & 5 & 15 \\ \hline
G32 & 1398 & 901 & 1390 & 221 & 1396 & 82345 & {\bf 1410} & 194 & 1402 & 17.5 &  & 5 & 5 \\ \hline
G33 & 1362 & 926 & 1360 & 198 & {\bf 1376} & 76282 & {\bf 1376} & 194 & 1374 & 15.9 &  & 1 & 1 \\ \hline
G34 & 1364 & 950 & 1368 & 237 & 1372 & 79406 & {\bf 1382} & 194 & 1374 & 15.9 &  & 24 & 24 \\ \hline
G35 & 7668 & 1258 & 7670 & 440 & 7635 & 167221 & 7485 & 263 & {\bf 7675} & 37.1 &  & 5 & 29 \\ \hline
G36 & 7660 & 1392 & 7660 & 400 & 7632 & 167203 & 7473 & 265 & {\bf 7663} & 37.6 &  & 3 & 58 \\ \hline
G37 & 7664 & 1387 & 7666 & 382 & 7643 & 170786 & 7484 & 288 & {\bf 7679} & 37.8 &  & 1 & 15 \\ \hline
G38 & {\bf 7681} & 1012 & 7646 & 1189 & 7602 & 178570 & 7479 & 264 & 7679 & 37.7 &  & 7 & 18 \\ \hline
G39 & 2393 & 1311 & 2395 & 852 & 2303 & 42584 & {\bf 2405} & 209 & 2404 & 37.2 &  & 1 & 1 \\ \hline
G40 & 2374 & 1166 & 2387 & 901 & 2302 & 39549 & 2378 & 208 & {\bf 2389} & 38.1 &  & 7 & 7 \\ \hline
G41 & 2386 & 1017 & 2398 & 942 & 2298 & 40025 & {\bf 2405} & 208 & 2401 & 37.8 &  & 20 & 71 \\ \hline
G42 & 2457 & 1458 & {\bf 2469} & 875 & 2390 & 41255 & 2465 & 210 & {\bf 2469} & 37.3 &  & 4 & 4 \\ \hline
G43 & 6656 & 406 & 6656 & 213 & 6659 & 35324 & 6658 & 245 & {\bf 6660} & 29.1 &  & 17 & 129 \\ \hline
G44 & {\bf 6648} & 356 & 6643 & 192 & 6642 & 34519 & 6646 & 241 & {\bf 6648} & 29.2 &  & 21 & 129 \\ \hline
G45 & 6642 & 354 & 6652 & 210 & 6646 & 34179 & 6652 & 241 & {\bf 6653} & 29.1 &  & 10 & 53 \\ \hline
G46 & 6634 & 498 & 6645 & 639 & 6630 & 38854 & 6647 & 245 & {\bf 6649} & 29.1 &  & 9 & 13 \\ \hline
G47 & 6649 & 359 & {\bf 6656} & 633 & 6640 & 36587 & 6652 & 242 & {\bf 6656} & 29.1 &  & 16 & 91 \\ \hline
G48 & {\bf 6000} & 20 & {\bf 6000} & 119 & {\bf 6000} & 64713 & {\bf 6000} & 210 & {\bf 6000} & 23.2 &  & 194 & 194 \\ \hline
G49 & {\bf 6000} & 35 & {\bf 6000} & 134 & {\bf 6000} & 64749 & {\bf 6000} & 210 & {\bf 6000} & 23.2 &  & 180 & 180 \\ \hline
G50 & {\bf 5880} & 27 & {\bf 5880} & 231 & {\bf 5880} & 147132 & 5858 & 211 & 5874 & 25.6 &  & 10 & 94 \\ \hline
G51 & {\bf 3846} & 513 & 3837 & 497 & 3808 & 89966 & 3841 & 234 & {\bf 3846} & 18.4 &  & 23 & 68 \\ \hline
G52 & {\bf 3849} & 551 & 3833 & 507 & 3816 & 95985 & 3845 & 228 & 3848 & 18.4 &  & 10 & 49 \\ \hline
G53 & {\bf 3846} & 424 & 3842 & 503 & 3802 & 92459 & 3845 & 230 & {\bf 3846} & 18.4 &  & 9 & 102 \\ \hline
G54 & 3846 & 429 & 3842 & 524 & 3820 & 98458 & 3845 & 228 & {\bf 3850} & 18.5 &  & 3 & 40 \\ \hline
        \end{tabular}
}
    \end{center}
  \caption{Results of oscillator-based Ising machines run on MAX-CUT benchmarks
in the G-set, compared with several heuristic algorithms. Time reported in this
table is for a single run. $n_{max}$ is the number of runs out of the 200
trials where the cut reaches the maximum of 200; $n_{0.999}$ is the number for
it to reach 99.9\% of the maximum.\tablabel{results}}

\end{table}

From \tabref{results}, we observe that our oscillator-based Ising machine is indeed effective --- it finds best-known cut values for 38 out of the 54 problems, 17 of which are even better than those reported in the above literature.
Moreover, in the 200 random instances, the best cut is often reached more than once --- the average $n_{max}$ for all benchmarks is 20 out of 200.
If we relax the objective and look at the number of instances where 99.9\% of the cut value is reached, represented by $n_{0.999}$, the average is 56, more than a quarter of the total trials.
The results can in fact be improved further if we tailor the annealing schedule for each problem.
But to show the effectiveness and generality of our scheme, we have chosen to use the same annealing schedule for all the problems. 

In the annealing schedule we used, the coupling strength $K$ increases linearly, the noise level $K_n$ steps up from 0 to 1, while SYNC's amplitude $K_s$ ramps up and down multiple times.
Such a schedule was chosen empirically and appears to work well for most G-set problems.
\figref{G1} shows the behaviour of oscillator phases and the instantaneous cut values under this schedule for solving benchmark problem G1 to its best-known cut size.
Some \MATLAB code to illustrate the annealing schedule is shown in \ref{app:code} %.
The code uses \MATLAB's SDE solver and is thus much slower than an implementation in C++ we used to generate the results in \tabref{results}. 
We plan to release all our code as open-source software in the summer of 2019 so that others can verify and build on our work.

\begin{figure}[htbp!]
    \centering
    {
        \epsfig{file=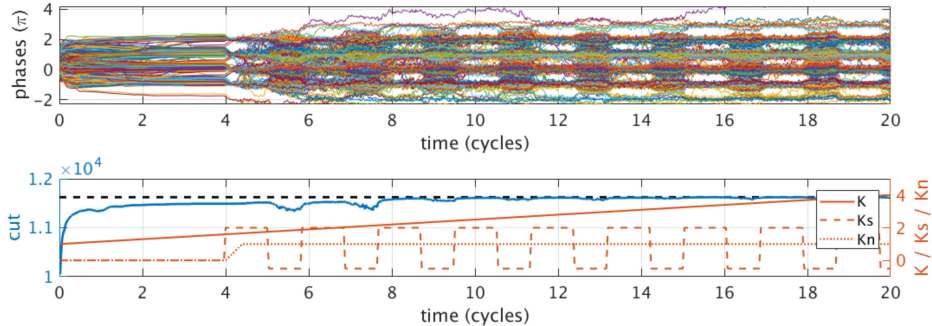, width=0.9\linewidth}
    }
	\caption{Coupled oscillators solving MAX-CUT benchmark problem G1 \cite{helmberg2000spectral}
             to its best-known cut size 11624.
    \figlabel{G1}}
\end{figure}

The fact that we were using a fixed schedule also indicates that the actual hardware time for the Ising machine to solve all these benchmarks is the same, regardless of problem size and connectivity.
Note that in \figref{G1}, the end time 20 means 20 oscillation cycles, but this end time is predicated on a coupling strength of $K\sim 1$.
The actual value of $K$ for each oscillator depends on the PPV function as well as the amplitude of perturbation from other oscillators, as we show in the derivation of Gen-Adler in \ref{app:derivation} %.
As an example, for the LC oscillators we use in \secref{smallMAXCUT} with 100k resistive coupling, $K\approx 0.02$.
This indicates that it takes less than 100 cycles for the oscillators to synchronise in phase, which is consistent with measurements.
For such a coupled LC oscillator network, a hardware time of 20 in \figref{G1} represents approximately 2000 cycles of oscillation; for 5MHz oscillators that takes 0.4ms.
If we use GHz nano-oscillators, the computation time can be well within a microsecond.
In comparison, the runtime of the several heuristic algorithms listed in \tabref{results}, even with faster CPUs and parallel implementations in the future, is unlikely to ever drop to this range.

As the hardware time is fixed, the runtime we report in \tabref{results} for our Ising machines is the time for simulating the SDEs of coupled oscillators on CPUs.
While we list runtime results for each algorithm in \tabref{results}, note that they come from different sources and are measured on different platforms.
Results for SS, CirCut and VNSPR were obtained from Dual Intel Xeon at 3.06GHz with 3.2GB of RAM; SA was run on Intel Xeon E3-1245v2 at 3.4GHz with 32GB of RAM \cite{myklebust2015SA}.
To make the results generally comparable, we ran our simulations on a modest personal desktop with Intel Xeon E5-1603v3 at 2.8GHz with 16GB of RAM.
Even so, it came as a nice surprise to us that even by simulating SDEs we were able to solve the benchmarks efficiently.
Another notable feature of our method is that unlike other algorithms, SDE simulation does not know about the Ising Hamiltonian or cut value --- it never needs to evaluate the energy function or relative energy changes, which are implicit in the dynamics of differential equations, yet it proves effective and fast.

\begin{figure}[htbp!]
    \centering
    {
        \epsfig{file=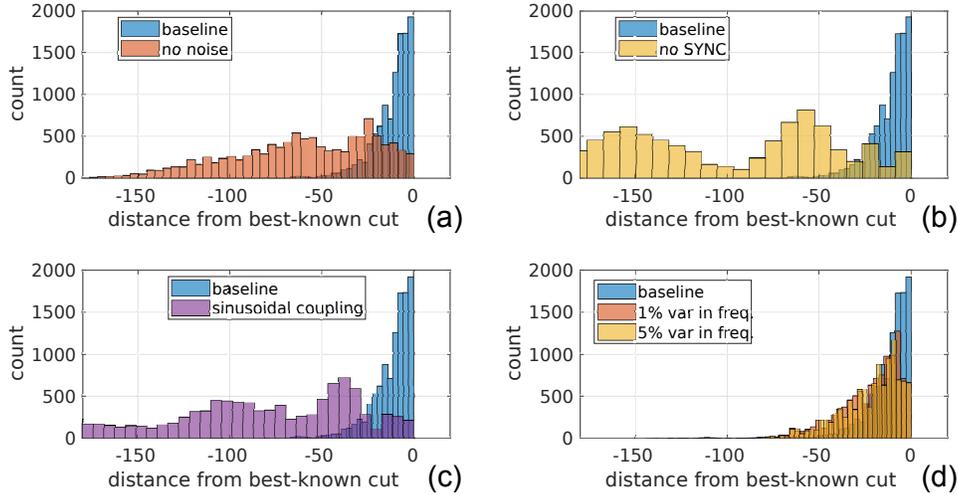, width=0.9\linewidth}
    }
    \vspace{0.5em}
	\caption{
        Histograms of the cut values achieved by several variants of the Ising machine, compared with the baseline results used in \tabref{results}.
        \figlabel{histograms}
    }
\end{figure}

We also ran more computational experiments on the G-set benchmarks in order to study the mechanism of oscillator-based Ising machines.
We created several variants of the Ising machine used above by removing different components in its operation.
For each variant, we re-ran 200 random instances for each of the 54 benchmarks, generating 10800 cut values.
In \figref{histograms}, we compare the quality of these cut values with results from the unaltered Ising machine by plotting histograms of the distances of the cut values to their respective maxima.
In the first variant, we removed noise from the model by setting $K_n\equiv 0$.  
The solutions become considerably worse, confirming that noise helps the coupled oscillator system settle to lower energy states.

In the next variant, we removed SYNC by setting $K_s\equiv 0$.
Without SYNC, the system becomes a simple coupled oscillator system with phases that take a continuum of values, as discussed in \secref{Kuramoto}.
The settled analog values of the phases that were then thresholded to $0$ or $\pi$ to correspond to Ising spins.
% Such a system is similar to the associative memory arrays proposed for
% specialised image processing tasks
% \cite{Hoppensteadt2000synchronization,maffezzoni2015oscArray,Porod2015physical}.
As shown in \figref{histograms}, the results become significantly worse; indeed, none of the best-known results were reached.
This indicates that the SYNC signal and the mechanism of SHIL we introduce to the coupled oscillator networks are indeed essential for them to operate as Ising machines.

Our baseline Ising machine actually uses a smoothed square function $\tanh(\sin(.))$ for the coupling, as opposed to the $\sin(.)$ used in the original Kuramoto model, as shown in the code in \ref{app:code} %.
This changes the $\cos(.)$ term in the energy function \eqnref{ESHIL1} to a triangle function.
Such a change appears to give better results than the original, as shown in \figref{histograms} (c).
The change requires designing oscillators with special PPV shapes and waveforms such that their cross-correlation is a square wave, which is not difficult in practice based on our derivation in \ref{app:derivation} %.
As an example, rotary traveling wave oscillators naturally have square PPVs.
Ring oscillators can also be designed with various PPVs and waveforms by sizing each stage individually.
We cannot say definitively that the square function we have used is optimal for Ising solution performance, but the significant improvement over sinusoidal coupling functions indicates that a fruitful direction for further exploration may be to look beyond the original Kuramoto model for oscillator-based computing.

The last variant we report here added variability to the natural frequencies of the oscillators, as in \eqnref{KuramotoVar}.
We assigned Gaussian random variables to $\omega_i$s with $\omega^*$ as the mean, and 0.01 (1\%) and 0.05 (5\%), respectively, as the standard deviations for two separate runs.
From \figref{histograms} (d), we observe that even with such non-trivial spread in the natural frequencies of oscillators, the performance is affected very little.

\begin{figure}[htbp!]
    \centering
    {
        \epsfig{file=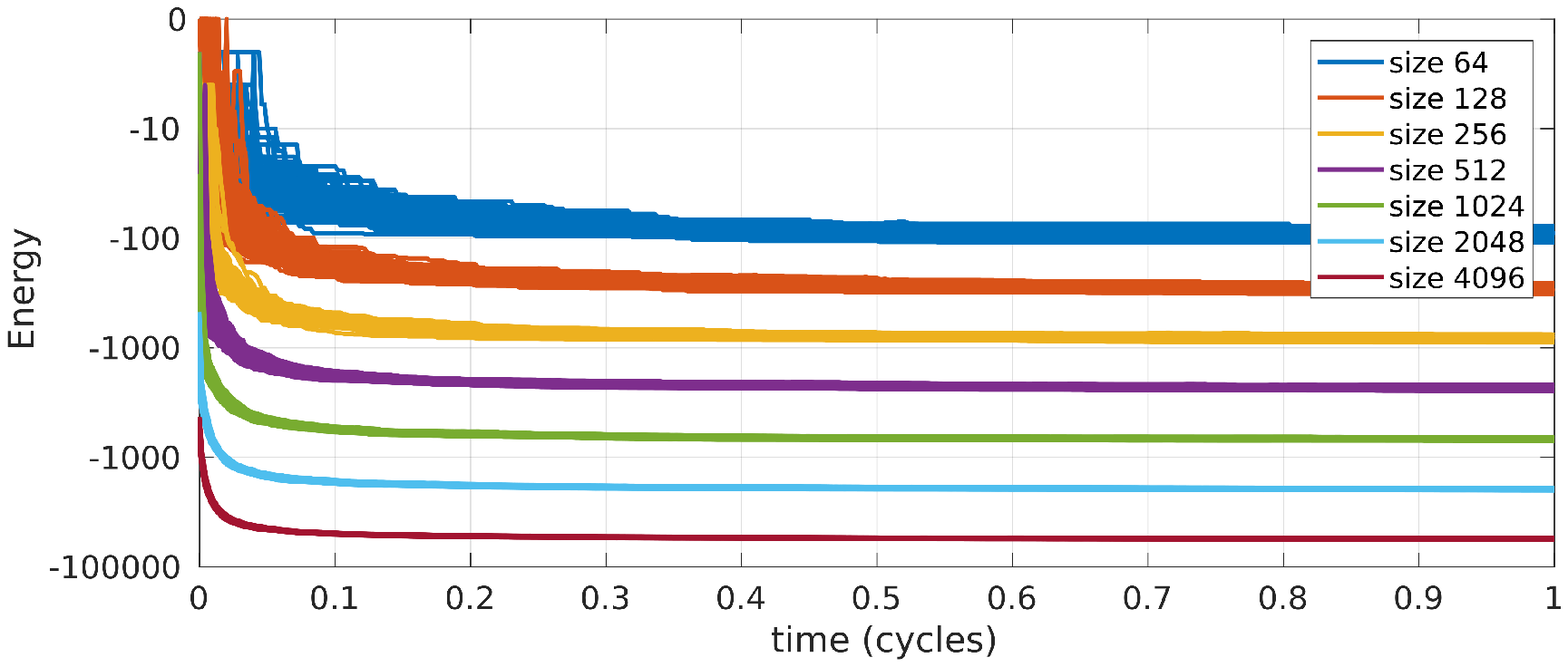, width=0.8\linewidth}
    }
    \caption{
    \figlabel{scaling_small}}
\end{figure}

% \subsection{Speed and Scaling}
Finally, we conducted a preliminary study of the scaling of the time taken by the Ising machine to reach good solutions as problem sizes increase.
As the G-set benchmarks have only a few sizes (800, 1000, 2000 and 3000), we used the program (named \texttt{rudy}) that generated them to create more problems of various sizes.
All generated problems used random graphs with 10\% connectivity and $\pm 1$ coupling coefficients.
We simulated all of them, each for 200 instances, with fixed parameters $K=1$, $K_s=0.1$, $K_n=0.01$, and show all their Ising Hamiltonians over time in \figref{scaling_small}.
Much to our surprise, the speed in which the values settle appears almost constant, regardless of the problem size.
While this does not necessarily mean they all converge to the global optima within the same time, this preliminary study is encouraging as it confirms the massively parallel nature of the system.
For larger Ising problems, our Ising machine only needs to scale linearly in hardware size with the number of spins, but does not necessarily require much more time to reach a solution.

\subsection{A Graph Colouring Example}

As mentioned in \secref{IsingIntro}, many problems other than MAX-CUT can be mapped to the Ising model \cite{lucas2013ising} and solved by an oscillator-based Ising machine.
Here we show an example of a graph colouring problem --- assigning four colours to the 51 states (including a federal district) of America such that no two adjacent states have the same colour.

Each state is represented as a vertex in the graph.
When two states are adjacent, there is an edge in the graph that connects the corresponding vertices.
For every vertex $i$, we assign four spins --- $s_{iR}$, $s_{iG}$, $s_{iB}$ and s$_{iY}$ to represent its colouring scheme; when only one of them is $+1$, the vertex is successfully coloured as either red, green, blue or yellow.
Then we write an energy function $H$ associated with these $4\times 51 = 204$ spins as follows:
\begin{align}\eqnlabel{colouring}
    H &= \sum_i^n (2 + s_{iR} + s_{iG} + s_{iB} + s_{iY})^2 \nonumber\\
    & + \sum_{(i,j)\in \mathbb{E}}^{n_\mathbb{E}} \left[(1+s_{iR})(1+s_{jR}) + (1+s_{iG})(1+s_{jG})
    + (1+s_{iB})(1+s_{jB}) + (1+s_{iY})(1+s_{jY}) \right],
\end{align}
where $n=51$ is the number of vertices, $\mathbb{E}$ represents the edge set, $n_\mathbb{E}$ is
the number of edges and in this case equal to 220.\footnote{Hawaii and Alaska are
considered adjacent such that their colours will be different in the map.}

The first term of $H$ is a sum of squares never less than zero; it reaches zero only when $\{s_{iR},~s_{iG},~s_{iB},~s_{iY}\}$ contains three $-1$s and one $+1$ for every $i$, \ie, each state has a unique colour.
The latter term is also a sum that is always greater than or equal to zero, as each spin can only take a value in $\{-1,~+1\}$; it is zero when $s_{iX} = s_{jX} = +1$ never occurs for any edge connecting $i$ and $j$, and for any colour $X \in \{R,~G,~B,~Y\}$, \ie, adjacent states do not share the same colour.
Therefore, when $H$ reaches its minimum value 0, the spin configuration represents a valid colouring scheme --- following the indices of the $+1$ spins $\{i,~X~|~s_{iX}=+1\}$, we can then assign colour $X$ to state $i$.

\begin{figure}[htbp!]
    \centering
    {
        \epsfig{file=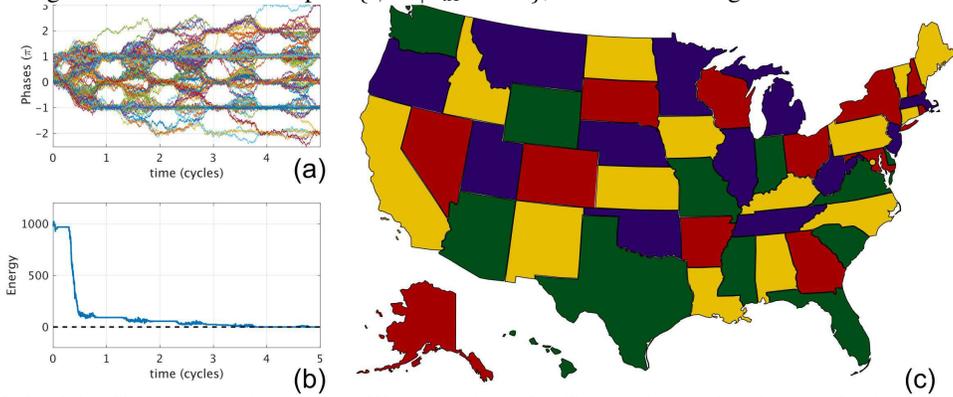, width=0.9\linewidth}
    }
    \caption{Coupled oscillators colouring the states in the US map:
             (a) phases of oscillators evolve over time;
             (b) energy function \eqnref{colouring} decreases during the process;
             (c) the resulting US map colouring scheme.
    \figlabel{US_map_with_phases_energy_small}}
\end{figure}

Note that when expanding the sum of squares in \eqnref{colouring}, we can use the fact $s_{iX}^2 \equiv 1$ to eliminate the square terms.
This means $H$ contains only products of two spins --- modelled by $J_{ij}$s, and self terms --- modelled by $h_{i}$.
These Ising coefficients can then be used to determine the couplings in an oscillator-based Ising machine.

We simulated these 204 coupled oscillators and show the results in \figref{US_map_with_phases_energy_small}.
In the simulation, we kept $K$ and $K_n$ constant while ramping $K_s$ up and down 5 times.
We found the Ising machine to be effective with this schedule as it could colour the map successfully in more than 50\% of the random trials and return many different valid colouring schemes.

\section*{{Conclusion}}
In this paper, we have proposed a novel scheme for implementing Ising machines using self-sustaining nonlinear oscillators.
We have shown how coupled oscillators naturally minimise an ``energy'' represented by their global Lyapunov function, and how introducing the mechanism of subharmonic injection locking modifies this function to encode the Ising Hamiltonian for minimisation.
The validity and feasibility of the scheme have been examined via multiple levels of simulation and proof-of-concept hardware implementations.
Simulations run on larger-scale benchmark problems have also shown promising results in both speed and the quality of solutions.
We believe that our scheme constitutes an important and practical means for the implementation of scalable Ising machines.

\renewcommand{\baselinestretch}{1.0}
\let\em=\it
{\scriptsize
\bibliographystyle{unsrt}
\bibliography{stringdefs,jr,von-Neumann-jr,PHLOGON-jr,tianshi}

\begin{thebibliography}{10}

\bibitem{ising1925beitrag}
{E. Ising}.
\newblock {Beitrag zur theorie des ferromagnetismus}.
\newblock {\em Zeitschrift f{\"u}r Physik A Hadrons and Nuclei},
  31(1):253--258, 1925.

\bibitem{Brush1967RevModPhysHistoryOfIsingModel}
Stephen~G. Brush.
\newblock {History of the Lenz-Ising Model}.
\newblock {\em Rev. Mod. Phys.}, 39:883--893, Oct 1967.

\bibitem{barahona1982computational}
Francisco Barahona.
\newblock {On the computational complexity of Ising spin glass models}.
\newblock {\em {Journal of Physics A: Mathematical and General}}, 15(10):3241,
  1982.

\bibitem{marandi2014network}
{A. Marandi, Z. Wang, K. Takata, R. L. Byer and Y. Yamamoto}.
\newblock {Network of time-multiplexed optical parametric oscillators as a
  coherent Ising machine}.
\newblock {\em Nature Photonics}, 8(12):937--942, 2014.

\bibitem{mcmahon2016ScienceIsing100}
{P. L. McMahon, A. Marandi, Y. Haribara, R. Hamerly, C. Langrock, S. Tamate and
  T. Inagaki, H. Takesue, S. Utsunomiya, K. Aihara and others}.
\newblock {A fully-programmable 100-spin coherent Ising machine with all-to-all
  connections}.
\newblock {\em Science}, page 5178, 2016.

\bibitem{inagaki2016ScienceIsing2000}
{T. Inagaki, Y. Haribara, K. Igarashi, T. Sonobe, S. Tamate, T. Honjo, A.
  Marandi, P. L. McMahon, T. Umeki, K. Enbutsu and others}.
\newblock {A Coherent Ising machine for 2000-node Optimization Problems}.
\newblock {\em Science}, 354(6312):603--606, 2016.

\bibitem{johnson2011quantum}
{M. W. Johnson, M. H. S. Amin, S. Gildert, T. Lanting, F. Hamze, N. Dickson, R.
  Harris, A. J. Berkley, J. Johansson, P. Bunyk and others}.
\newblock {Quantum Annealing with Manufactured Spins}.
\newblock {\em Nature}, 473(7346):194, 2011.

\bibitem{bian2014Ising}
{Z. Bian, F. Chudak, R. Israel, B. Lackey, W. G. Macready and A. Roy}.
\newblock {Discrete optimization using quantum annealing on sparse Ising
  models}.
\newblock {\em Frontiers in Physics}, 2:56, 2014.

\bibitem{yamaoka2016IsingCMOS}
{M. Yamaoka, C. Yoshimura, M. Hayashi, T. Okuyama, H. Aoki and H. Mizuno}.
\newblock {A 20k-spin Ising Chip to Solve Combinatorial Optimization Problems
  with CMOS Annealing}.
\newblock {\em IEEE Journal of Solid-State Circuits}, 51(1):303--309, 2016.

\bibitem{karp1972np}
{R. M. Karp}.
\newblock {Reducibility among combinatorial problems}.
\newblock In {\em Complexity of computer computations}, pages 85--103.
  Springer, 1972.

\bibitem{lucas2013ising}
{A. Lucas}.
\newblock {Ising formulations of many NP problems}.
\newblock {\em arXiv preprint arXiv:1302.5843}, 2013.

\bibitem{festa2002randomized}
{P. Festa, P. M. Pardalos, M. G. C. Resende, and C. C. Ribeiro}.
\newblock {Randomized heuristics for the MAX-CUT problem}.
\newblock {\em Optimization methods and software}, 17(6):1033--1058, 2002.

\bibitem{Jensen2011GraphColoringProblems}
Tommy~R Jensen and Bjarne Toft.
\newblock {\em {Graph Coloring Problems}}, volume~39.
\newblock {John Wiley \& Sons}, 2011.

\bibitem{NeRoDATE2012SHIL}
{A. Neogy and J. Roychowdhury}.
\newblock {Analysis and Design of Sub-harmonically Injection Locked
  Oscillators}.
\newblock In {\em Proc. IEEE DATE}, Mar 2012.
\newblock \putDOI{\href{http://dx.doi.org/10.1109/DATE.2012.6176677}{DOI
  link.}}

\bibitem{BhRoASPDAC2009}
P.~Bhansali and J.~Roychowdhury.
\newblock {Gen-Adler: The generalized Adler's equation for injection locking
  analysis in oscillators}.
\newblock In {\em Proc. IEEE ASP-DAC}, pages 522--227, January 2009.
\newblock \href{http://dx.doi.org/10.1109/ASPDAC.2009.4796533}{DOI link.}

\bibitem{Kuramoto1975}
Yoshiki Kuramoto.
\newblock Self-entrainment of a population of coupled non-linear oscillators.
\newblock In {\em International symposium on mathematical problems in
  theoretical physics}, pages 420--422. Springer, 1975.

\bibitem{Kuramoto2003}
{Y. Kuramoto}.
\newblock {\em {Chemical Oscillations, Waves and Turbulence}}.
\newblock Dover, 2003.

\bibitem{acebron2005kuramoto}
{J. A. Acebr{\'o}n, L. L. Bonilla, C. J. P. Vicente, F. Ritort and R. Spigler}.
\newblock {The Kuramoto Model: A Simple Paradigm for Synchronization
  Phenomena}.
\newblock {\em {Reviews of Modern Physics}}, 77(1):137, 2005.

\bibitem{WaRoUCNC2014PHLOGON}
{T. Wang and J. Roychowdhury}.
\newblock {PHLOGON: PHase-based LOGic using Oscillatory Nanosystems}.
\newblock In {\em Proc. UCNC}, LNCS sublibrary: Theoretical computer science
  and general issues. Springer, July 2014.
\newblock \putDOI{\href{http://dx.doi.org/10.1007/978-3-319-08123-6_29}{DOI
  link.}}

\bibitem{AramonEtAlarXiv2018DigitalAnnealer}
{Maliheh Aramon, Gili Rosenberg, Elisabetta Valiante, Toshiyuki Miyazawa,
  Hirotaka Tamura, and Helmut G. Katzgraber}.
\newblock {Physics-Inspired Optimization for Quadratic Unconstrained Problems
  Using a Digital Annealer}.
\newblock In {\em {arXiv:1806.08815 [physics.comp-ph]}}, August 2018.

\bibitem{GyHiSaIEICE2018isingNoNoise}
{H.~Gyoten, M.~Hiromoto and T.~Sato}.
\newblock Area efficient annealing processor for ising model without random
  number generator.
\newblock {\em {IEICE Transactions on Information and Systems}},
  E101.D(2):314--323, 2018.

\bibitem{GyHiSaICCAD2018parallelTemperingForIsing}
{H.~Gyoten, M.~Hiromoto and T.~Sato}.
\newblock Enhancing the solution quality of hardware ising-model solver via
  parallel tempering.
\newblock In {\em Proc. ICCAD}, ICCAD '18, pages 70:1--70:8, New York, NY, USA,
  2018. ACM.

\bibitem{bian2010ising}
Zhengbing Bian, Fabian Chudak, William~G Macready, and Geordie Rose.
\newblock {The Ising model: teaching an old problem new tricks}.
\newblock {\em {D-Wave Systems}}, 2, 2010.

\bibitem{harris2010flux}
{R. Harris, J. Johansson, A. J. Berkley, M. W. Johnson, T. Lanting, S. Han, P.
  Bunyk, E. Ladizinsky, T. Oh, I. Perminov and others}.
\newblock {Experimental Demonstration of a Robust and Scalable Flux Qubit}.
\newblock {\em {Physical Review B}}, 81(13):134510, 2010.

\bibitem{ronnow2014defining}
{T. F. R{\o}nnow, Z. Wang, J. Job, S. Boixo, S. V. Isakov, D. Wecker, J. M.
  Martinis, D. A. Lidar and M. Troyer}.
\newblock {Defining and detecting quantum speedup}.
\newblock {\em Science}, 345(6195):420--424, 2014.

\bibitem{denchev2016tunneling}
{V. S. Denchev, S. Boixo, S. V. Isakov, N. Ding, R. Babbush, V. Smelyanskiy, J.
  Martinis, H. Neven}.
\newblock {What is the computational value of finite-range tunneling?}
\newblock {\em Physical Review X}, 6(3):031015, 2016.

\bibitem{mahboob2016electromechanical}
{I. Mahboob, H. Okamoto and H. Yamaguchi}.
\newblock {An electromechanical Ising Hamiltonian}.
\newblock {\em Science Advances}, 2(6):e1600236, 2016.

\bibitem{camsari2017pbits}
{K. Y. Camsari, R. Faria, B. M. Sutton and S. Datta}.
\newblock {Stochastic p-Bits for Invertible Logic}.
\newblock {\em Physical Review X}, 7(3):031014, 2017.

\bibitem{yamamoto2017time}
{K. Yamamoto, W. Huang, S. Takamaeda-Yamazaki, M. Ikebe, T. Asai, M. Motomura}.
\newblock {A Time-Division Multiplexing Ising Machine on FPGAs}.
\newblock In {\em Proceedings of the 8th International Symposium on Highly
  Efficient Accelerators and Reconfigurable Technologies}, page~3. ACM, 2017.

\bibitem{Winfree67}
A.~Winfree.
\newblock {Biological Rhythms and the Behavior of Populations of Coupled
  Oscillators}.
\newblock {\em Theoretical Biology}, 16:15--42, 1967.

\bibitem{DeMeRoTCAS2000}
A.~Demir, A.~Mehrotra, and J.~Roychowdhury.
\newblock {Phase Noise in Oscillators: a Unifying Theory and Numerical Methods
  for Characterization}.
\newblock {\em IEEE Trans.~Ckts.~Syst.~--~I: Fund.~Th.~Appl.}, 47:655--674, May
  2000.
\newblock \putDOI{\href{http://dx.doi.org/10.1109/81.847872}{DOI link.}}

\bibitem{WaRoDAC2015MAPPforPHLOGON}
{T. Wang and J. Roychowdhury}.
\newblock {Design Tools for Oscillator-based Computing Systems}.
\newblock In {\em Proc.\/ IEEE DAC}, pages 188:1--188:6, 2015.
\newblock \putDOI{\href{http://dx.doi.org/10.1145/2744769.2744818}{DOI link.}}

\bibitem{shinomoto1986phase}
{S. Shinomoto, Y. Kuramoto}.
\newblock {Phase transitions in active rotator systems}.
\newblock {\em Progress of Theoretical Physics}, 75(5):1105--1110, 1986.

\bibitem{lyapunov1992general}
A.~M. Lyapunov.
\newblock {The General Problem of the Stability of Motion}.
\newblock {\em International Journal of Control}, 55(3):531--534, 1992.

\bibitem{myklebust2015SA}
{T. Myklebust}.
\newblock {Solving maximum cut problems by simulated annealing}.
\newblock {\em arXiv preprint arXiv:1505.03068}, 2015.

\bibitem{helmberg2000spectral}
{C. Helmberg, F. Rendl}.
\newblock {A spectral bundle method for semidefinite programming}.
\newblock {\em SIAM Journal on Optimization}, 10(3):673--696, 2000.

\bibitem{marti2009SS}
{R. Mart{\'\i}, A. Duarte, M. Laguna}.
\newblock {Advanced scatter search for the max-cut problem}.
\newblock {\em INFORMS Journal on Computing}, 21(1):26--38, 2009.

\bibitem{burer2002circut}
{S. Burer, R. Monteiro, Y. Zhang}.
\newblock {Rank-two relaxation heuristics for max-cut and other binary
  quadratic programs}.
\newblock {\em SIAM Journal on Optimization}, 12(2):503--521, 2002.

\bibitem{Adler46}
R.~Adler.
\newblock A study of locking phenomena in oscillators.
\newblock {\em Proceedings of the I.R.E. and Waves and Electrons}, 34:351--357,
  June 1946.

\bibitem{Roychowdhury05}
J.~Roychowdhury.
\newblock Exact analytical equations for predicting nonlinear phase errors and
  jitter in ring oscillators.
\newblock In {\em Proc. IEEE International Conference on VLSI Design}, January
  2005.

\bibitem{DeRoTCAD2003}
A.~Demir and J.~Roychowdhury.
\newblock {A Reliable and Efficient Procedure for Oscillator PPV Computation,
  with Phase Noise Macromodelling Applications}.
\newblock {\em IEEE Trans. CAD}, pages 188--197, February 2003.
\newblock \putDOI{\href{http://dx.doi.org/10.1109/TCAD.2002.806599}{DOI link.}}

\bibitem{MeRoICCAD2006PPVHB}
{T. Mei and J. Roychowdhury}.
\newblock {PPV-HB: Harmonic Balance for Oscillator/PLL Phase Macromodels}.
\newblock In {\em Proc. ICCAD}, pages 283--288, Nov. 2006.

\bibitem{Ro2001BookChap1}
J.~Roychowdhury.
\newblock {\em Applied and Computational Control, Signals and Circuits --
  Recent Developments}, chapter 3 (Multi-time PDEs for Dynamical System
  Analysis), pages 85--143.
\newblock Kluwer Academic, 2001.

\bibitem{Adler1973}
R.~Adler.
\newblock {A study of locking phenomena in oscillators}.
\newblock {\em Proc. IEEE}, 61:1380--1385, 1973.
\newblock Reprinted from \cite{Adler46}.

\bibitem{landau1969statistical}
{L. D. Landau and E. M. Lifshitz}.
\newblock {\em {Statistical Physics: V. 5: Course of Theoretical Physics}}.
\newblock Pergamon press, 1969.

\bibitem{yuan2010lyapunov}
{R. Yuan, Y. Ma, B. Yuan and P. Ao}.
\newblock {Constructive Proof of Global Lyapunov Function as Potential
  Function}.
\newblock {\em arXiv preprint arXiv:1012.2721}, 2010.

\end{thebibliography}
}

% \begin{appendices}
\appendix
\renewcommand{\thesection}{\appendixname~\Alph{section}.}

\clearpage
\raggedbottom % llncs seems to align page bottoms for appendices
\section{Phase-based Macromodels of Oscillators and Oscillator-based Ising Machines}\applabel{derivation}

Studies on the phase dynamics of coupled oscillators commonly start from the
Kuramoto model or its variants, which are high-level abstractions.
Here in this appendix, we start from low-level oscillator models instead and
derive the generalised Kuramoto model for Ising machines from the ground up.

A single nonlinear self-sustaining oscillator is an autonomous dynamical system,
usually modelled using Differential Algebraic Equations (DAEs) in the following
form.
\begin{equation}\eqnlabel{DAEnoB}
	\frac{d}{dt} \vec q(\vec x(t)) + \vec f(\vec x(t))  = \vec 0,
\end{equation}
where $\vec x(t) \in \mathbf{R}^n$ represents the unknowns in the system;
$\vec q(.)$ and $\vec f(.)$ are $\colon \mathbf{R}^n \to \mathbf{R}^n$ functions,
representing the differential and algebraic parts of the DAE respectively.

A self-sustaining oscillator has a nonconstant $T_0$-periodic solution $\vec
x^*_s(t)$ to \eqnref{DAEnoB}, satisfying $\vec x^*_s(t+T_0) = \vec x^*_s(t)$.

When this oscillator is under a time-varying perturbation modelled as $\vec
b(t) \in \mathbf{R}^n$, its DAE can be written as follows.
\begin{equation}\eqnlabel{DAE}
	\frac{d}{dt} \vec q(\vec x(t)) + \vec f(\vec x(t))  + \vec b(t) = \vec 0.
\end{equation}

If the perturbation is small, the oscillator's perturbed response can be
approximated well as
\begin{equation}\eqnlabel{xt}
	\vec x^*(t) = \vec x^*_s(t + \alpha(t)),
\end{equation}
where $\alpha(t)$ is the phase shift caused by the external input and is
governed by the following differential equation \cite{DeMeRoTCAS2000}:
\begin{equation}\eqnlabel{PPV}
	\frac{d}{dt} \alpha(t) = \vec p^T(t + \alpha(t)) \cdot \vec b(t)
\end{equation}
where the time-varying vector $\vec p(t)$ is known as the Perturbation
Projection Vector (PPV) \cite{DeMeRoTCAS2000} of the oscillator.
It is $T_0$-periodic, and is a property intrinsic to the oscillator that
captures its phase response to small external inputs.
$\vec p(t)$ can be derived analytically \cite{Roychowdhury05}
or calculated numerically \cite{DeRoTCAD2003,MeRoICCAD2006PPVHB} from
the oscillator's DAE without knowing any information about the input $\vec
b(t)$.

To study injection locking in the oscillator, we assume that the external input
$\vec b(t)$ is a periodic signal with frequency
$\omega^* = 2\pi f^* = 2\pi /T^*$ and a time-varying phase $\phi_{in}(t)$:
\begin{equation}\eqnlabel{PPVphi}
\vec b(t) =  \vec b_{(2\pi)}(\omega^*\cdot t + \phi_{in}(t)),
\end{equation}
where $\vec b_{(2\pi)}$ is $2\pi$-periodic function.

Equation \eqnref{PPV} can be rewritten as
\begin{equation}\eqnlabel{PPVphiTWO}
\frac{d}{dt} \phi(t) = \omega_0-\omega^* + \omega_0 \cdot \vec p_{(2\pi)}^T(\omega^*\cdot t +
\phi(t)) \cdot \vec b_{(2\pi)}(\omega^*\cdot t + \phi_{in}(t)),
\end{equation}
where $\phi(t) = (\omega_0-\omega^*)\cdot t + \omega_0\cdot \alpha(t)$
--- when injection locking occurs, it is the phase shift of the oscillator's
response from a perfect oscillation at $\omega^*$.
$\vec p_{(2\pi)}$ is the $2\pi$-periodic version of the PPV ---
$\vec p_{(2\pi)}(t) = \vec p(t/\omega_0)$.

A differential equation such as \eqnref{PPVphi} can be formulated as
a Multi-time Partial Differential Equation (MPDE) \cite{Ro2001BookChap1}, assuming the oscillation
at $\omega^*$ happens in the fast time $t_2$ whereas the phases evolve in the
slow time $t_1$.
\begin{align}\eqnlabel{PPVphiMPDE}
\frac{\partial \hat\phi(t_1,~t_2)}{\partial t_1} + \frac{\partial \hat\phi(t_1,~t_2)}{\partial t_2}
= & ~\omega_0-\omega^* \nonumber\\
& + \omega_0 \cdot \vec p_{(2\pi)}^T(\omega^*\cdot t_2 +
\hat\phi(t_1)) \cdot \vec b_{(2\pi)}(\omega^*\cdot t_2 + \phi_{in}(t_1)).
\end{align}

The theory of MPDE states that if we solve the PDE in \eqnref{PPVphiMPDE} for 
$\hat\phi^*(.,~.)$, the solution to the original single-time differential
equation \eqnref{PPVphi} can be written as $\phi^*(t) = \hat\phi^*(t,~t)$.

We can approximate the MPDE by averaging it along the $t_2$ dimension ---
replacing the $t_2$-varying solution with a constant for every $t_1$.
This is usually a valid approximation that does not degrade the accuracy by
much, as the phase of oscillation is changing much more slowly than the
oscillation itself.
In this case, we are approximating the two-dimensional solution
$\hat\phi(t_1,~t_2)$ with $\bar\phi(t_1)$.

\begin{align}\eqnlabel{PPVphiMPDEt1}
\frac{d \bar\phi(t_1)}{d t_1}
= & ~\omega_0-\omega^* \nonumber\\
& + \omega_0 \cdot \int_0^{2\pi} \vec p_{(2\pi)}^T(\omega^*\cdot t_2 +
\bar\phi(t_1)) \cdot \vec b_{(2\pi)}(\omega^*\cdot t_2 + \phi_{in}(t_1)) ~ d t_2 \\
= & ~\omega_0-\omega^* \nonumber\\
& + \omega_0 \cdot \int_0^{2\pi} \vec p_{(2\pi)}^T(\omega^*\cdot t_2 +
\bar\phi(t_1) -  \phi_{in}(t_1)) \cdot \vec b_{(2\pi)}(\omega^*\cdot t_2) ~ d t_2.
\end{align}

To help simplify the equation, we define
\begin{equation}
c(t) = \int_0^{2\pi} \vec p_{(2\pi)}^T(t + \tau) \cdot \vec b_{(2\pi)}(\tau) ~ d \tau.
\end{equation}

$c(t)$ is a $2\pi$-periodic function.
It is the cross-correlation\footnote{It is also known as the sliding dot
product or sliding inner product.} of the two functions $\vec p_{(2\pi)}(.)$
and $\vec b_{(2\pi)}(.)$.

The one-dimensional solution $\bar\phi(t_1)$ to \eqnref{PPVphiMPDEt1} can then
be used as an approximation for the solution of $\phi(t)$.
Put in other words, we can rewrite the differential equation \eqnref{PPVphi}
for the phase dynamics as
\begin{equation}\eqnlabel{GenAdler}
\frac{d}{d t} \phi(t) = ~\omega_0-\omega^* + \omega_0 \cdot c(\phi(t) -  \phi_{in}(t)).
\end{equation}

% \begin{figure}[htb]
%     \centering
%     {
%         \epsfig{file=figures/Screenshot_Gen_Adler.eps, height=0.4\linewidth,width=0.7\linewidth}
%     }
%     \caption{Placeholder figure from Gen-Adler paper. Remake it to illustrate averaging over fast time.
%     \figlabel{Screenshot_Gen_Adler}}
% \end{figure}

In the case of a common LC oscillator, both its waveform and PPV are sinusoidal.
With resistive coupling, its PPV $\vec p_{(2\pi)}(.)$ is proportional to
$\cos(t)$ \cite{Adler1973}.
When the perturbation is the same as the oscillator's waveform, \ie,
$\vec b_{(2\pi)}(t) = V\cdot \sin(t)$, the resulting $c(.)$ function is
proportional to $\sin(.)$.
Then we have rederived the Adler's equation for LC oscillators.
\begin{equation}
\frac{d}{d t} \phi(t) = ~\omega_0-\omega^* + \omega_0 \cdot A \cdot \sin(\phi(t) -  \phi_{in}(t)),
\end{equation}
where the coupling strength $A$ is determined mainly by the $Q$ factor of the
LC oscillator.

Another special case is when $\vec b(t)$ is a second-order perturbation, \ie,
it is oscillating at $2\omega^*$.
We can define another $2\pi$-periodic function $\vec b^{(2)}_{2\pi}(t)$,
such that $\vec b_{2\pi}(t) = \vec b^{(2)}_{2\pi}(2t)$.
If we further assume that the PPV is also oscillating at twice its natural
frequency, \ie, $\vec p_{2\pi}(t) = \vec p^{(2)}_{2\pi}(2t)$,
it can be proven that $c(t)$ is also a second-order oscillation.
\begin{align}
 c(t) &= \int_0^{2\pi} \vec p^{(2)T}_{(2\pi)}(2t + 2\tau) \cdot \vec b^{(2)}_{(2\pi)}(2\tau) ~ d \tau \\
      &= \frac{1}{2}\int_0^{4\pi} \vec p^{(2)T}_{(2\pi)}(2t + \tau) \cdot \vec b^{(2)}_{(2\pi)}(\tau) ~ d \tau \\
      &= \int_0^{2\pi} \vec p^{(2)}_{(2\pi)T}(2t + \tau) \cdot \vec b^{(2)}_{(2\pi)}(\tau) ~ d \tau \\
      &\triangleq c^{(2)}(2t),
\end{align}
where $c^{(2)}(.)$ is a $2\pi$-periodic function, making $c(.)$ is $\pi$-periodic.

In this special case, the phase dynamics \eqnref{GenAdler} can then be written as
\begin{equation}
\frac{d}{d t} \phi(t) = ~\omega_0-\omega^* + \omega_0 \cdot c^{(2)}(2(\phi(t) -  \phi_{in}(t))).
\end{equation}

For the $i$th oscillator in an oscillator-based Ising machine, the perturbation
to its phase comes from several sources: its connections to the other
oscillators, its connection to a reference signal for implementing the self
terms in the Ising Hamiltonian, and the second-order SYNC signal.
From the above discussion, we can write its phase dynamics as
\begin{align}\eqnlabel{GenKuramoto}
\frac{d}{dt} \phi_i(t) = & ~\omega_i-\omega^* + \omega_i \cdot K \cdot \sum_{j=1,~j\neq i}^n \left[J_{ij} \cdot \vec p_{ij}^T(\omega^*\cdot t +
\phi_i(t)) \cdot \vec b_j(\omega^*\cdot t + \phi_{j}(t))\right] \nonumber\\
& + \omega_i \cdot K\cdot h_i \cdot \vec p_i^T(\omega^*\cdot t + \phi_i(t)) \cdot \vec b_0(\omega^*\cdot t)  \nonumber\\
& + \omega_i \cdot K_s\cdot \vec p_{is}^T(\omega^*\cdot t + \phi_i(t)) \cdot \vec b_s(\omega^*\cdot t).
\end{align}

In equation \eqnref{GenKuramoto}, $\omega_i$ is the frequency of this oscillator,
$\omega^*$ is the central frequency for the coupled oscillators.
$\vec p_{ij}(.)$ is the $2\pi$-periodic PPV of the $i$th oscillator, with perturbation from the $j$th oscillator,
$\vec b_j(.)$ is the $2\pi$-periodic perturbation from the $j$th oscillator.
$\vec p_i(.)$ is the $2\pi$-periodic PPV of the $i$th oscillator, with perturbation from a reference signal,
$\vec b_0(.)$ is the $2\pi$-periodic reference signal.  $\vec b_s(.)$ and $\vec
p_{is}(.)$ represent the waveform of SYNC and the PPV entries corresponding to
this second-order perturbation respectively, and are thus both $\pi$-periodic.
We can define $\vec b_s(t) = \vec b^{(2)}_s(2t)$, 
$\vec p_{is}(t) = \vec p^{(2)}_{is}(2t)$.

Furthermore, we define
\begin{align}
c_{ij}(t) =& \int_0^{2\pi} \vec p_{ij}^T(t + \tau) \cdot \vec b_j(\tau) ~ d \tau.\\
d_i(t) =& \int_0^{2\pi} \vec p_i^T(t + \tau) \cdot \vec b_0(\tau) ~ d \tau.\\
s_i(t) =& \int_0^{2\pi} \vec p^{(2)T}_{is}(t + \tau) \cdot \vec b^{(2)}_s(\tau) ~ d \tau.
\end{align}

Then \eqnref{GenKuramoto} can be approximated well with
\begin{align}\eqnlabel{KuramotoFinal}
\frac{d}{d t} \phi_i(t) =& ~\omega_i-\omega^* + \omega_i \cdot K \cdot
\sum_{j=1,~j\neq i}^n \left[ J_{ij} \cdot c_{ij}(\phi_i(t) -  \phi_j(t)) \right]
\nonumber\\
& + \omega_i \cdot K \cdot h_i \cdot \vec d_i(\phi_i(t)) + \omega_i \cdot K_s \cdot \vec s_i(2\phi_i(t))
\end{align}

We can then write the global Lyapunov function as
\begin{equation}\eqnlabel{LyapunovFinal}
	E(\vec \phi(t)) = K\left[\sum_{i, j} J_{ij} \cdot C_{ij}(\phi_i(t) - \phi_j(t)) + 
	\sum_{i=1}^n h_i \cdot D_{i}(\phi_i(t))\right] + 
	\sum_{i=1}^n K_s \cdot S_{i}\left(2\phi_i(t)\right)
      - 2\sum_{i=1}^n \frac{\omega_i - \omega^*}{\omega_i}\phi_i,
\end{equation}
where $C_{ij}(t)$, $D_{i}(t)$ and $S_{i}(t)$ are defined as follows.

\begin{align}
C_{ij}(t) =& \int_0^{t} c_{ij}(\tau) ~ d \tau + C_{0ij}.\\
D_{i}(t) =& \int_0^{t} d_{i}(\tau) ~ d \tau + D_{0i}.\\
S_{i}(t) =& \int_0^{t} s_{i}(\tau) ~ d \tau + S_{0i}.\\
\end{align}
where $C_{0ij}$, $D_{0i}$ and $S_{0i}$ are arbitrary constants.

The generalised Kuramoto model we use in \secref{main} is just a special case
for the phase macromodel we present in this appendix for coupled oscillators.
When we need $C_{ij}$ to be $-\cos(.)$ functions, we choose $c_{ij}$ to be
$\sin(.)$ and use the PPV of harmonic oscillators.
In fact, based on the theory we develop in this section, we have all the
flexibility to change the shape of periodic functions $C_{ij}$ and $D_{i}$.
Oscillators can be designed or tweaked to yield the desired energy function.

\ignore{
In the remainder of this appendix, we give the proof that $E(\vec \phi(t))$ as
defined in \eqnref{LyapunovFinal} is indeed a global Lyapunov function of
\eqnref{KuramotoFinal}.

\textit{Proof:} Using the chain rule of differentiation, we have
\be{chainRuleOnL}
    \ddt{E}{t} = \sum_{l=1}^n \ddx{E}{\phi_l} \, \ddt{\phi_l}{t}.
\ee
Consider $\ddx{E}{\phi_l}$. Using \er{LyapunovFinal}, we have
\be{pLpThetalSplit}
    \begin{split}
        \ddx{E}{\phi_l} = &
            \overbrace{\ddx{}{\phi_l}\left\{ \sum_{i=1}^n \sum_{k=1}^n - \frac{1}{n} \left[ \Delta \omega_i \, \phi_i + \Delta \omega_k \, \phi_k \right] \right\}}^A
            + \overbrace{\ddx{}{\phi_l}\left\{ \sum_{i=1}^n \sum_{k=1}^n \frac{1}{2n} \left[ S_i(2\phi_i) + S_k(2\phi_k) \right] \right\}}^B \\
            & + \overbrace{\ddx{}{\phi_l}\left\{ \sum_{i=1}^n \sum_{k=1}^n J_{ik} \, C_{ik}(\phi_i - \phi_k)\right\}}^C.
    \end{split}
\ee
We consider the three terms $A$, $B$ and $C$ in \er{pLpThetalSplit} separately, starting with $C$.
\be{termCderiv1}
    \begin{split}
    C & \triangleq \ddx{}{\phi_l}\left\{ \sum_{i=1}^n \sum_{k=1}^n J_{ik} \, C_{ik}(\phi_i - \phi_k) \right\}
      = \sum_{i=1}^n \sum_{k=1}^n J_{ik} \, \ddx{}{\phi_l} C_{ik}(\phi_i - \phi_k) \\
      & = - \sum_{i=1}^n \sum_{k=1}^n \left[ J_{ik} \, (\delta_{il} - \delta_{kl}) \, z(\phi_i - \phi_k) \right] \quad \text{(using \er{Idefn}),}
    \end{split}
\ee
where
\be{deltailDefn}
    \delta_{il} \triangleq \begin{cases} 1 & \text{if } i = l, \\ 0 & \text{otherwise.}\end{cases}
\ee
Hence 
\be{termCderiv2}
    \begin{split}
    C = & - \left[ \sum_{k=1}^n \sum_{i=1}^n \delta_{il} J_{ik}  \, z(\phi_i - \phi_k) 
        - \sum_{i=1}^n \sum_{k=1}^n \delta_{kl} J_{ik} \, z(\phi_i - \phi_k) \right] \\
      = &- \left[ \sum_{k=1}^n J_{lk}  \, z(\phi_l - \phi_k) 
        - \sum_{i=1}^n  J_{il} \, z(\phi_i - \phi_l) \right] \\
      = &- \left[ \sum_{k=1}^n J_{lk}  \, z(\phi_l - \phi_k) 
        + \sum_{k=1}^n  J_{lk} \, z(\phi_l - \phi_k) \right] \quad (\text{using \er{zxAssumption} and $J_{il}=J_{li}$})\\
      = & - 2 \sum_{k=1}^n  J_{lk}\, z(\phi_l - \phi_k).
    \end{split}
\ee
Turning to $B$, we have
\be{termBderiv}
    \begin{split}
    B \triangleq & \ddx{}{\phi_l}\left\{ \sum_{i=1}^n \sum_{k=1}^n \frac{1}{2n} \left[ S_i(2\phi_i) + S_k(2\phi_k) \right] \right\}
      = \sum_{i=1}^n \sum_{k=1}^n \frac{1}{2n} \left[ \ddx{}{\phi_l} S_i(2\phi_i) + \ddx{}{\phi_l} S_k(2\phi_k) \right] \\
      = & - \frac{1}{2n} \sum_{i=1}^n \sum_{k=1}^n \left[2 \delta_{il} \, z_s(2\phi_i) + 2 \delta_{kl} \, z_s(2\phi_k) \right] \quad \text{(using \er{Isdefn})}\\
      = & -\frac{1}{n} \sum_{k=1}^n \sum_{i=1}^n \delta_{il} \, z_s(2\phi_i) - \frac{1}{n} \sum_{i=1}^n \sum_{k=1}^n \delta_{kl} \, z_s(2\phi_k) 
        = -\frac{1}{n} \sum_{k=1}^n z_s(2\phi_l) -+ \frac{1}{n} \sum_{i=1}^n z_s(2\phi_l) \\
      = & -2 z_s(2\phi_l).
    \end{split}
\ee
Finally, for $A$, we have
\be{termAderiv}
    \begin{split}
    A \triangleq & \ddx{}{\phi_l}\left\{ \sum_{i=1}^n \sum_{k=1}^n - \frac{1}{n} \left[ \Delta \omega_i \, \phi_i + \Delta \omega_k \, \phi_k \right] \right\}
      = -\frac{1}{n} \sum_{i=1}^n \sum_{k=1}^n \left[ \Delta \omega_i \, \delta_{il} + \Delta \omega_k \, \delta_{kl} \right]  \\
      = &-\frac{1}{n} \sum_{k=1}^n \sum_{i=1}^n \Delta \omega_i \, \delta_{il}  - \frac{1}{n} \sum_{i=1}^n \sum_{k=1}^n \Delta \omega_k \, \delta_{kl}
        = -\frac{1}{n} \sum_{k=1}^n \Delta \omega_l - \frac{1}{n} \sum_{i=1}^n \Delta \omega_l \\
      = & - 2 \Delta \omega_l.
    \end{split}
\ee 
Using \er{termAderiv}, \er{termBderiv} and \er{termCderiv2}, \er{pLpThetalSplit} can be expressed as
\be{pLpThetal}
    \begin{split}
    \ddx{L}{\phi_l} = & -2 \left[ \Delta \omega_l + z_s(2 \phi_l) + \sum_{k=1}^n J+{lk} \, z(\phi_l- \phi_k) \right]. \\
                      = & -2 \frac{1}{f_{0l}} \ddt{\phi_l(t)}{t} \quad \text{(using \er{KSwSHIL})}.
    \end{split}
\ee
Using \er{pLpThetal}, \er{chainRuleOnL} becomes
\be{dLdtFinal}
    \ddt{L}{t} = \sum_{l=1}^n \ddx{L}{\phi_l} \, \ddt{\phi_l}{t} = -2 \sum_{l=1}^n \frac{1}{f_{0l}} \left(\ddx{L}{\phi_l}\right)^2 \le 0.
\ee
Thus, \eqnref{LyapunovFinal} is indeed a global Lyapunov function of the system modelled in \eqnref{KuramotoFinal}.  $\halmos$

}

\section{Stochastic Model of Oscillator-based Ising Machines}\applabel{noise}

Noise in the phases of oscillators is commonly modelled by adding white noise
sources to the oscillator frequencies:
\begin{equation}\eqnlabel{KuramotoNoise}
	\frac{d}{dt} \phi_i(t) = - K\cdot \sum_{j=1,~j\neq i}^n J_{ij} \cdot \sin(\phi_i(t) - \phi_j(t))
						- K_s \cdot \sin(2\phi_i(t)) + K_n\cdot\xi_i(t),
\end{equation}
where variable $\xi_i(t)$ represents Gaussian white noise with a zero mean and
a correlator $\langle\xi_i(t),~\xi_i(\tau)\rangle = \delta(t-\tau)$;
scaler $K_n$ represents the magnitude of noise.

\eqnref{KuramotoNoise} can be rewritten as a stochastic differential equation (SDE).
\begin{equation}\eqnlabel{KuramotoSDE}
	d\phi_{it} = \left[ - K\cdot \sum_{j=1,~j\neq i}^n J_{ij} \cdot \sin(\phi_{it} - \phi_{jt})
						- K_s \cdot \sin(2\phi_{it}) \right]dt + K_n\cdot dW_t,
\end{equation}
and can then be simulated with standard SDE solvers.

To analyse the steady states of this SDE, we can apply the Boltzmann law from
statistical mechanics \cite{landau1969statistical}.
For a system with discrete states $\vec s_i$, $i=1,\cdots,M$, if each state is
associated with an energy $E_i$,\footnote{It is provable that a global Lyapunov function, 
if it exists, can be used instead of a physical energy to derive the same Boltzmann law
\cite{yuan2010lyapunov}.}
the probability $P_i$ for the system to be at each state can be written as
follows.
\begin{equation}
	P_i = {\frac{e^{- E_i / k T}}{\sum_{j=1}^{M}{e^{- E_j / k T}}}},
\end{equation}
where $k$ is the Boltzmann constant, $T$ is the thermodynamic temperature of
the system.
While $k$ and $T$ are concepts specific to statistical mechanics, in this
context the product $kT$ corresponds to the noise level $K_n$.

Given two spin configurations $\vec s_1$ and $\vec s_2$, the ratio between
their probabilities is known as the Boltzmann factor:
\begin{equation}
    \frac{P_2}{P_1} = e^{\frac{E_1 - E_2}{kT}}.
\end{equation}

According to the energy function \eqnref{ESHIL2} associated with
oscillator-based Ising machines, the energy difference that determines this
factor is proportional to the coupling strength.
\begin{equation}
    E_1 - E_2 \propto K.
\end{equation}

If $\vec s_1$ is the higher energy state, \ie, $E_1 > E_2$, as the coupling
strength $K$ increases, it becomes less and less likely for the system to
stay at $\vec s_1$.
The system prefers the lowest energy state in the presence of noise.

\section{Proof for the global Lyapunov function of oscillator-based Ising machines}
\applabel{proof}

Our starting point is the Gen-Adler \ignore{(Kuramoto-Sakaguchi)} equation under 2-SHIL (which we have derived elsewhere):
\be{KSwSHIL}
\frac{1}{\omega_{0l}} \ddt{\phi_l(t)}{t} = \underbrace{\frac{\omega_{0l} - \frac{\omega_s}{2}}{\omega_{0l}}}_{\Delta \omega_l} + z_s(2 \phi_l(t)) + \sum_{k=1}^N J_{ik} \, z(\phi_l(t)-\phi_k(t)), \quad l=1, \cdots, N,
\ee
where $N$ is the number of coupled oscillators in the system,  $\omega_{0l}$ is the natural frequency of the $l^\text{th}$ oscillator, $\phi_l(t)$ is the phase of the $l^\text{th}$ oscillator, $\omega_s \simeq 2 \omega_{0l}$ is the frequency of the SYNC signal, $z_s(\cdot)$ is  the $2\pi$-periodic Adlerized function for the SYNC input to each oscillator, $J_{lk}$ is the symmetric coupling between the $l^\text{th}$ and $k^\text{th}$ oscillators, and $z(\cdot)$ is the $2\pi$-periodic Adlerized function for the coupling inputs to each oscillator.

We assume that every oscillator in the coupled system is sub-harmonically locked to the frequency $\frac{\omega_s}{2}$ (more precisely, that $\phi_l(t)$ remain bounded and small $\forall t, \forall l$), and that $z(x)$ is skew-symmetric, \ie, 
\be{zxAssumption}
    z(-x) = - z(x).
\ee

Define
\be{Idefn}
    I(x) \triangleq - \int_0^x z(y) \, dy + K,
\ee
and
\be{Isdefn}
    I_s(x) \triangleq -\int_0^x z_s(y) \, dy + K_s,
\ee
where $K$ and $K_s$ are arbitrary constants.

\begin{theorem}
\thmlabel{dLdtLTEzero}
\be{Ldefn}
    L\big(\phi_1,\cdots,\phi_N\big) \triangleq \sum_{i=1}^N \sum_{k=1}^N 
    \left\{
        - \frac{1}{N} \left[ \Delta \omega_i \phi_i + \Delta \omega_k \phi_k \right]
        + \frac{1}{2N} \left[ I_s(2\phi_i) + I_s(2\phi_k) \right]
        + J_{ik} \, I(\phi_i - \phi_k)
    \right\}
\ee
constitutes a Lyapunov function for \er{KSwSHIL}, \ie, 
\be{dLdtLTE0}
    \ddt{L\big(\phi_1(t), \cdots, \phi_N(t)\big)}{t} \le 0 \quad \forall t.
\ee
\end{theorem}
\textit{Proof:} Using the chain rule of differentiation, we have
\be{chainRuleOnL}
    \ddt{L}{t} = \sum_{l=1}^N \ddx{L}{\phi_l} \, \ddt{\phi_l}{t}.
\ee
Consider $\ddx{L}{\phi_l}$. Using \er{Ldefn}, we have
\be{pLpThetalSplit}
    \begin{split}
        \ddx{L}{\phi_l} = &
            \overbrace{\ddx{}{\phi_l}\left\{ \sum_{i=1}^N \sum_{k=1}^N - \frac{1}{N} \left[ \Delta \omega_i \, \phi_i + \Delta \omega_k \, \phi_k \right] \right\}}^A
            + \overbrace{\ddx{}{\phi_l}\left\{ \sum_{i=1}^N \sum_{k=1}^N \frac{1}{2N} \left[ I_s(2\phi_i) + I_s(2\phi_k) \right] \right\}}^B \\
            & + \overbrace{\ddx{}{\phi_l}\left\{ \sum_{i=1}^N \sum_{k=1}^N J_{ik} \, I(\phi_i - \phi_k)\right\}}^C.
    \end{split}
\ee
We consider the three terms $A$, $B$ and $C$ in \er{pLpThetalSplit} separately, starting with $C$.
\be{termCderiv1}
    \begin{split}
    C & \triangleq \ddx{}{\phi_l}\left\{ \sum_{i=1}^N \sum_{k=1}^N J_{ik} \, I(\phi_i - \phi_k) \right\}
      = \sum_{i=1}^N \sum_{k=1}^N J_{ik} \, \ddx{}{\phi_l} I(\phi_i - \phi_k) \\
      & = - \sum_{i=1}^N \sum_{k=1}^N \left[ J_{ik} \, (\delta_{il} - \delta_{kl}) \, z(\phi_i - \phi_k) \right] \quad \text{(using \er{Idefn}),}
    \end{split}
\ee
where
\be{deltailDefn}
    \delta_{il} \triangleq \begin{cases} 1 & \text{if } i = l, \\ 0 & \text{otherwise.}\end{cases}
\ee
Hence 
\be{termCderiv2}
    \begin{split}
    C = & - \left[ \sum_{k=1}^N \sum_{i=1}^N \delta_{il} J_{ik}  \, z(\phi_i - \phi_k) 
        - \sum_{i=1}^N \sum_{k=1}^N \delta_{kl} J_{ik} \, z(\phi_i - \phi_k) \right] \\
      = &- \left[ \sum_{k=1}^N J_{lk}  \, z(\phi_l - \phi_k) 
        - \sum_{i=1}^N  J_{il} \, z(\phi_i - \phi_l) \right] \\
      = &- \left[ \sum_{k=1}^N J_{lk}  \, z(\phi_l - \phi_k) 
        + \sum_{k=1}^N  J_{lk} \, z(\phi_l - \phi_k) \right] \quad (\text{using \er{zxAssumption} and $J_{il}=J_{li}$})\\
      = & - 2 \sum_{k=1}^N  J_{lk}\, z(\phi_l - \phi_k).
    \end{split}
\ee
Turning to $B$, we have
\be{termBderiv}
    \begin{split}
    B \triangleq & \ddx{}{\phi_l}\left\{ \sum_{i=1}^N \sum_{k=1}^N \frac{1}{2N} \left[ I_s(2\phi_i) + I_s(2\phi_k) \right] \right\}
      = \sum_{i=1}^N \sum_{k=1}^N \frac{1}{2N} \left[ \ddx{}{\phi_l} I_s(2\phi_i) + \ddx{}{\phi_l} I_s(2\phi_k) \right] \\
      = & - \frac{1}{2N} \sum_{i=1}^N \sum_{k=1}^N \left[2 \delta_{il} \, z_s(2\phi_i) + 2 \delta_{kl} \, z_s(2\phi_k) \right] \quad \text{(using \er{Isdefn})}\\
      = & -\frac{1}{N} \sum_{k=1}^N \sum_{i=1}^N \delta_{il} \, z_s(2\phi_i) - \frac{1}{N} \sum_{i=1}^N \sum_{k=1}^N \delta_{kl} \, z_s(2\phi_k) 
        = -\frac{1}{N} \sum_{k=1}^N z_s(2\phi_l) -+ \frac{1}{N} \sum_{i=1}^N z_s(2\phi_l) \\
      = & -2 z_s(2\phi_l).
    \end{split}
\ee
Finally, for $A$, we have
\be{termAderiv}
    \begin{split}
    A \triangleq & \ddx{}{\phi_l}\left\{ \sum_{i=1}^N \sum_{k=1}^N - \frac{1}{N} \left[ \Delta \omega_i \, \phi_i + \Delta \omega_k \, \phi_k \right] \right\}
      = -\frac{1}{N} \sum_{i=1}^N \sum_{k=1}^N \left[ \Delta \omega_i \, \delta_{il} + \Delta \omega_k \, \delta_{kl} \right]  \\
      = &-\frac{1}{N} \sum_{k=1}^N \sum_{i=1}^N \Delta \omega_i \, \delta_{il}  - \frac{1}{N} \sum_{i=1}^N \sum_{k=1}^N \Delta \omega_k \, \delta_{kl}
        = -\frac{1}{N} \sum_{k=1}^N \Delta \omega_l - \frac{1}{N} \sum_{i=1}^N \Delta \omega_l \\
      = & - 2 \Delta \omega_l.
    \end{split}
\ee 
Using \er{termAderiv}, \er{termBderiv} and \er{termCderiv2}, \er{pLpThetalSplit} can be expressed as
\be{pLpThetal}
    \begin{split}
    \ddx{L}{\phi_l} = & -2 \left[ \Delta \omega_l + z_s(2 \phi_l) + \sum_{k=1}^N J+{lk} \, z(\phi_l- \phi_k) \right]. \\
                      = & -2 \frac{1}{\omega_{0l}} \ddt{\phi_l(t)}{t} \quad \text{(using \er{KSwSHIL})}.
    \end{split}
\ee
Using \er{pLpThetal}, \er{chainRuleOnL} becomes
\be{dLdtFinal}
    \ddt{L}{t} = \sum_{l=1}^N \ddx{L}{\phi_l} \, \ddt{\phi_l}{t} = -2 \sum_{l=1}^N \frac{1}{\omega_{0l}} \left(\ddx{L}{\phi_l}\right)^2 \le 0.
\ee
Thus, \thmref{dLdtLTEzero} stands proved. $\halmos$

\ignore{
If we assume that
\be{zsForKuramoto}
    z(x) = z_s(x) \triangleq sin(2 \pi x),
\ee
corresponding to the basic Kuramoto model \cite{Kuramoto2003}, then  \er{Idefn} and \er{Isdefn} become (choosing $K=K_s=\frac{1}{2\pi}$)
\be{IsForKuramoto} 
    I(x) = I_s(x) = \frac{1}{2 \pi} \cos(2 \pi x),
\ee
and the Lyapunov function \er{Ldefn} becomes
\be{KuramotoLyapunov}
    \begin{split}
    L\big(\phi_1,\cdots,\phi_N\big) = \sum_{i=1}^N \sum_{k=1}^N 
    \left\{
        - \frac{1}{N} \left[ \Delta \omega_i \phi_i + \Delta \omega_k \phi_k \right] \right.
        & + \frac{1}{2N\cdot 2\pi} \left[ \cos(2 \pi \, 2\phi_i) + \cos (2 \pi \, 2\phi_k) \right] \\
        & \left. + \frac{1}{2\pi} J_{ik} \, \cos(2\pi(\phi_i - \phi_k)) \right\}.
    \end{split}
\ee
}

\section{\MATLAB SDE Simulation Code for MAX-CUT Problems}\applabel{code}

\matlabscript{code/KuramotoF_sin.m}{\texttt{KuramotoF\_sin.m}}
\matlabscript{code/run_MAXCUT_8.m}{\texttt{run\_MAXCUT\_8.m}}

\matlabscript{code/KuramotoF.m}{\texttt{KuramotoF.m}}
\matlabscript{code/run_MAXCUT_G1.m}{\texttt{run\_MAXCUT\_G1.m}}

% \end{appendices}

\end{document}